\documentclass[twocolumn]{aastex701}

\usepackage{amsmath}
\usepackage{color}
\usepackage{orcidlink}
\usepackage{soul}
\usepackage{array}

\definecolor{forestgreen}{RGB}{34,139,34}
\definecolor{maroon}{RGB}{139,0,0}

\newcounter{qnumber}

\begin{document}

\title{Searching for Free-Floating Planets with TESS: Results from Sectors 61 -- 65} 

\correspondingauthor{Michelle Kunimoto}
\email{mkuni@phas.ubc.ca}

\author[0000-0001-9269-8060]{Michelle Kunimoto}
\email{mkuni@phas.ubc.ca}
\affiliation{Department of Physics and Astronomy, University of British Columbia, 6224 Agricultural Road, Vancouver, BC V6T 1Z1, Canada}
\affiliation{Department of Physics and Kavli Institute for Astrophysics and Space Research, Massachusetts Institute of Technology, 77 Massachusetts Avenue, Cambridge, MA 02139, USA}

\author[0000-0003-1827-9399]{William DeRocco}
\affiliation{Maryland Center for Fundamental Physics, University of Maryland, College Park, 4296 Stadium Drive, College Park, MD 20742, USA}
\affiliation{Department of Physics \& Astronomy, The Johns Hopkins University, 3400 N. Charles Street, Baltimore, MD 21218, USA}
\email{derocco@umd.edu}

\author[0000-0002-8454-3015]{Nolan Smyth}
\affiliation{Department of Physics, Universit{\'e} de Montr{\'e}al, 1375 Avenue Th{\'e}r{\`e}se-Lavoie-Roux, Montr{\'e}al, QC H2V 0B3, Canada}
\affiliation{Ciela Institute, Universit{\'e} de Montr{\'e}al, 1375 Avenue Th{\'e}r{\`e}se-Lavoie-Roux, Montr{\'e}al, QC H2V 0B3, Canada}
\affiliation{Mila - Quebec Artificial Intelligence Institute, Montr{\'e}al QC H2S 3H1, Canada}
\email{nolanwsmyth@gmail.com}

\author[0000-0003-0081-1797]{Steve Bryson}
\affiliation{NASA Ames Research Center, Moffett Field, CA 94035, USA}
\email{steve.bryson@nasa.gov}

\author[0000-0003-0395-9869]{B. Scott Gaudi}
\affiliation{Department of Astronomy, The Ohio State University, Columbus, OH 43210, USA}
\email{gaudi.1@osu.edu }



\begin{abstract}
Though free-floating planets (FFPs) may outpopulate their bound counterparts in the terrestrial-mass range, they remain one of the least explored exoplanet demographics. Due to their negligible electromagnetic emission at all wavelengths, the only observational technique able to detect these worlds is gravitational microlensing. Microlensing by terrestrial-mass FFPs induces rare, short-duration magnifications of background stars, requiring high-cadence, wide-field surveys to detect these events. The Transiting Exoplanet Survey Satellite (TESS), though designed to detect close-bound exoplanets via transits, boasts a Full-Frame Image cadence as short as 200 seconds and has monitored hundreds of millions of stars, providing a unique dataset in which to search for rare short-duration transients. We have performed a preliminary search for FFP microlensing in 7.5 million light curves from TESS Sectors 61--65. We find one short-duration event with a light curve morphology consistent with expectations for a low-mass FFP, but in tension with the expected FFP abundance in this mass range. We consider possible false positive interpretations of this event such as stellar flares, hearbeat binaries, and centrifugal breakout. We find that all interpretations pose some challenges, and discuss the possibility that the event may constitute a first example of a new class of pernicious false positives that future space-based microlensing efforts will encounter. Our ongoing search through the TESS dataset will significantly support the upcoming hunt for rogue worlds with dedicated space-based microlensing surveys, and our results may be used alongside these surveys to place interesting constraints on the spatial distribution of FFPs in the Galaxy.

\end{abstract}


\section{Introduction} \label{sec:intro}

Free-floating planets (FFPs), planets not bound to any star, constitute an enigmatic planetary demographic of which little is known. Theory and simulation suggest that such planets should be ubiquitous in the Galaxy \citep{Strigari_2012, 2018ApJ...852...85H}, with different formation mechanisms dominating at different FFP masses. At high masses ($\gtrsim M_{\text{Jup}}$), they may form in isolation from the collapse of gas, constituting the extreme low-mass end of the stellar mass function \citep{Padoan_2002, Bonnell_2008, zwart2024origin}. FFPs formed through these processes may retain sufficient heat to be detected in the infrared, and recent observations have discovered an unexpectedly large abundance of such worlds \citep{2023arXiv231003552M, 2023arXiv231001231P}. At terrestrial masses, however, the dominant formation mechanism is thought to be gravitational ejection from a parent system during the chaotic early phases of system formation, e.g. through planet-planet scattering \citep{1996Sci...274..954R, 1996Natur.384..619W, Veras_2012}, ejection by an inner binary \citep{2003MNRAS.345..233N, Sutherland_2016, Smullen_2016, coleman2023constraining, Standing_2023, chen2023tilted, coleman2024properties}, or stellar fly-bys \citep{wang2024floating, yu2024freefloating}.

Such terrestrial-mass FFPs likely outpopulate their bound counterparts \citep{sumi2023freefloating, mroz2023exoplanet}, comprising the \textit{majority} of Galactic exoplanets in this mass range. However, their negligible emission across the electromagnetic spectrum poses a significant observational challenge. The only technique sensitive to detecting these worlds is gravitational microlensing, in which the gravitational field of the FFP (``lens'') focuses light from a distant background star (``source''), resulting in the temporary apparent magnification of the star as the lens traverses the source along the line of sight. Ground-based microlensing searches have recently yielded the first three observations of terrestrial-mass FFPs \citep{Mr_z_2019, koshimoto2023terrestrial, Mr_z_2020}. Though these observations have confirmed the existence of this population and suggested a high Galactic abundance, ground-based observations are limited by their photometric sensitivity and cadence, with atmospheric interference and technical requirements requiring magnifications of $\gtrsim 10\%$ and a cadence of $>15$ min \cite[e.g.,][]{Mr_z_2019}. Typical microlensing events for terrestrial-mass FFPs have timescales on the order of an hour, and can have relatively low peak magnifications, making their detection challenging for existing ground-based observatories.

Multiple dedicated space-based microlensing surveys are set to launch in the coming years, opening up a new era in the search for free-floating planets. However, with the increase in sensitivity that these missions will achieve brings new challenges. Chief among them is mitigating potential new classes of false positive that have not been well-characterized by ground-based efforts. The Transiting Exoplanet Survey Satellite \cite[TESS;][]{TESS} has provided a relatively unexplored dataset from which new short duration signals can be discovered. While TESS was optimized for detecting exoplanets via the transit method, its wide-field imaging capabilities and high-cadence, high-precision photometric observations make it sensitive to detecting brightening features similar to those expected for short-duration microlensing events as well. TESS observes large $24^{\circ}\times96^{\circ}$ sectors of the sky for 27.4 days at a time, allowing it to monitor millions of stars in its Full-Frame Images (FFIs). TESS's high photometric precision allows it to detect flux changes as small as a fraction of a percent. Furthermore, its current 200-second FFI cadence enables sensitivity to events lasting significantly less than an hour. While the proximity of most TESS target stars to the Sun results in a low a priori probability of any individual star being lensed by a foreground object over the duration of TESS observations, the large number of stars monitored by TESS over many years, combined with the potentially large population of terrestrial-mass FFP planet in our galaxy inferred from ground-based microlensing surveys, implies that the probability of detecting one or more FFP microlensing event in the extant TESS dataset is non-negligible, as we show Sec. \ref{sec:yield} and was also shown by \citet{yang2024rare}.

In this work, we leveraged a subset of the large catalog of existing TESS observations to perform a preliminary search for FFPs within TESS Sectors 61 -- 65. We used a dataset consisting  of 7.5 million stars with magnitudes as dim as $T = 15$ mag observed at a 200-second cadence. This initial search resulted in one potential short-duration microlensing event; however, though the light curve of this event is consistent with that of a terrestrial-mass FFP, we show that the event rate implied by this detection is in tension with expectations for this mass range, making it an unlikely scenario. We therefore examine alternative interpretations of the event, such as a flare, and find that these non-FFP interpretations are also unlikely; however, the limited observational data for such flares leaves us unable to make a quantitative statement about the nature of the event.
This underscores the need for potential false positives to be better characterized prior to future space-based microlensing surveys such as the Nancy Grace Roman Space Telescope \citep[Roman;][]{Akeson2019,Penny2019} and Earth 2.0 \citep{Ge2022}. This is a task that existing TESS observations are well-suited for.

\section{TESS Search for Free-Floating Planets} \label{sec:search}

\subsection{Microlensing Fundamentals}

When a foreground FFP crosses near the line of sight to a background source star, its gravitational field perturbs the emitted light rays and creates multiple images of the source. When these multiple images can not be individually resolved, the net effect is an apparent time-varying magnification of the source known as gravitational microlensing. The characteristic angular size of the lens, the \textit{Einstein angle}, is given by
\begin{equation}
\label{eq:RE}
    \theta_E = \sqrt{\frac{4 G M  (1 - d_{l}/d_{s})}{d_{l} c^2}},
\end{equation}
where $M$ is the mass of the lens and $d_{l}$ ($d_{s}$) is the distance to the lens (source).

The time it takes to traverse the Einstein radius is therefore
\begin{equation}
\label{eq:tE}
    t_E = \frac{\theta_E}{\mu_{\text{rel}}},
\end{equation}
where $\mu_{\text{rel}}$ is the proper motion of the source-lens system relative to the observer. The net magnification of a point-source during a microlensing event can be described analytically by an overall multiplicative factor relative to its baseline flux via \citep{1986ApJ...304....1P}
\begin{equation}
    A(t) = \frac{u^2 + 2}{u \sqrt{u^2 + 4}},
\end{equation}
where $u$ is the \textit{impact parameter}, the displacement of the lens from the source in units of the Einstein radius. For an event centered around time $t_0$, corresponding to a minimum value of $u_0$, the impact parameter evolves as 
\begin{equation}
    u = \sqrt{\Big(\frac{t - t_0}{t_E} \Big)^2 + u_0^2}.
\end{equation}
This allows one to model the magnification of the source in terms of $u_0, t_E,$ and $ t_0$. Along with these three parameters, two additional parameters, the baseline flux of the source ($f_s$) and blended background ($f_b$), provide a full parametrization of a point-source point-lens microlensing event.

When the angular extent of the source $\theta_{s}$ is comparable to or greater than $\theta_E$, the additional parameter $\rho = \theta_{s}/\theta_E$ describes the extent of finite-source effects in the light-curve. Finite-source effects generically reduce the peak magnification and elongate the duration of the event \cite[see, e.g.,][]{2009ApJ...695..200L}, resulting in a box-like morphology that makes the light-curve amenable to initial flagging through methods such as the Box Least Squares (BLS) search algorithm (see \S\ref{sec:detection} and \S\ref{sec:modeling}).

\subsection{Expected Event Yield}
\label{sec:yield}

Microlensing is an intrinsically rare phenomenon. Microlensing surveys rely upon the observation of a large number of source stars at a rapid cadence over a long observational baseline. TESS's transit survey was not designed with microlensing in mind; its ability to detect such events is limited with respect to dedicated microlensing campaigns due to its short per-sector observational baselines, relatively nearby stars, and choice of target fields. However, over the course of its Primary and Extended Missions, it has observed hundreds of millions of stars with continuous baselines ranging from 27.4 days to 1 year at cadences as rapid as 200 seconds. Though its targets are generally much closer and do not lie in the crowded stellar fields favored by dedicated microlensing searches, TESS's observations still provide the opportunity to search for short-duration events that may point to a large abundance of Galactic FFPs. Due to TESS's short observational baseline, any putative detection would need to be cross-referenced with long-baseline ground-based observations in order to better characterize such a signal. Beyond characterizing short-duration false positives for upcoming missions such as Roman and Earth 2.0, there also remains a small possibility of identifying an event that, coupled with other observations, could be identified as an FFP. As such, we performed estimates of the detection rate for such signals. As stated previously, TESS is not designed as a microlensing survey, and therefore the potential detection rate does  not compete with those of upcoming missions such as  Roman and Earth 2.0; however, as we will show below, even in the case of a null observation in the full TESS dataset, TESS's existing observations have the potential to help place interesting constraints on the largely uncertain abundance of FFPs in the sub-terrestrial range. Such a constraint would allow more refined predictions of the expected number of detections that Roman and Earth 2.0 may yield, providing further motivation for our ongoing search. Furthermore, by virtue of the proximity of TESS's targets, TESS data could provide a measurement of or constraint on local FFP abundance, which would prove valuable in characterizing the Galactic distribution of FFPs observed by upcoming microlensing surveys.

\begin{figure}
        \centering
        \includegraphics[width=\linewidth]{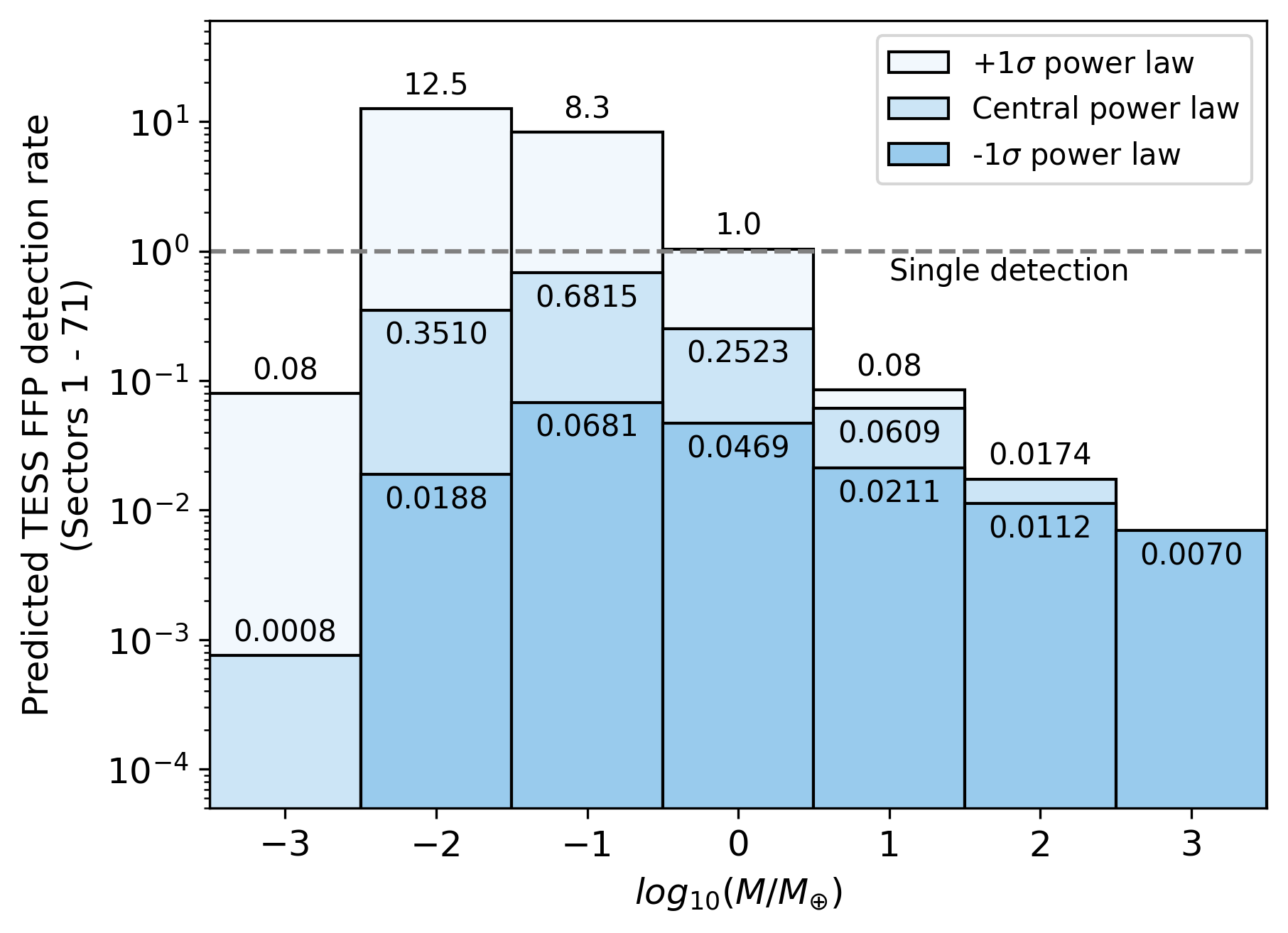}
        \caption{Expected FFP detection rate per dex in mass, across all currently available TESS sectors from QLP (Sectors 1 -- 71). The range of values correspond to different underlying mass functions, based on the central best-fit values describing the Sumi et al. power law and the $\pm1\sigma$ uncertainties \citep{sumi2023freefloating}. TESS is most sensitive to FFP events between $0.01 < M/M_{\oplus} < 0.1$. 
        }
        \label{fig:yields}
        \vspace{0.3in}
    \end{figure}

We estimate the yield of FFPs in TESS data following the methodology used by \citet{yang2024rare}. Our expected yields are shown in Figure \ref{fig:yields}, where each bin displays the expected number of detectable FFPs per dex in mass after searching 67 million TESS QLP light curves ($T < 13.5~\mathrm{mag}$) across Sectors 1 -- 71. The yields were computed based on assuming a signal-to-noise ratio (SNR) threshold of SNR $> 10$ and requiring at least five in-event datapoints. Note that the definition of SNR we employ (which we motivate in \S\ref{sec:detection}) differs from that of \citet{yang2024rare}, and results in marginally higher yield estimates than theirs. We adopt the FFP mass function derived by \citet{sumi2023freefloating} using data from existing ground-based observations, and display both the yield assuming the best-fit values for the model as well as the yields for the one standard deviation uncertainties on the fit parameters. 
    
At the best-fit values, we predict $\approx 1$ FFP within the TESS dataset, consistent with the prediction of \citet{yang2024rare}, while at the $+1\sigma$ level we predict as many as $\approx 20$ FFPs. These predictions highlight the present uncertainty on the overall abundance of FFPs in the sub-terrestrial mass range and motivates our ongoing work with TESS, which, even in the case of a null detection, would place improved constraints on the largely-uncertain low-mass FFP abundance.

\section{Methodology}

\subsection{Light Curves}
\label{sec:lightcurves}

We used FFI light curves extracted by the Quick-Look Pipeline \cite[QLP;][]{QLP1}, which covers all stars in the TESS Input Catalog \cite[currently TICv8.2;][]{TIC82} brighter than $T = 13.5$ mag. Since Sector 41, QLP also produces light curves for M dwarfs as faint as $T = 15$ mag \citep{QLP2}. For TESS Sectors 61 -- 65 (2023 January 18 -- June 1), we obtained light curves with 200-second cadence for a total of 7,499,939 targets.

From each light curve, we removed all data points flagged as poor quality as indicated by a quality flag of 1, as well as all data within 0.2 days of the start and end of each TESS orbit, resulting in an average observational baseline per target of 25.0 days. These times correspond to high-amplitude systematics caused by scattered light from the Earth or Moon. To remove low-frequency trends caused by stellar variability, we applied the biweight detrending algorithm implemented in \texttt{wotan} \citep{wotan} using a detrending window of 2 days. Given that we expect most terrestrial FFP events to have typical durations of less than one day, the 2-day window was chosen to minimize the impact of event distortion caused by detrending while removing long-term astrophysical and systematic instrumental trends.

\subsection{Microlensing Event Detection}
\label{sec:detection}

Transiting exoplanets are commonly found using the BLS period search algorithm \citep{BLS}, which searches for periodic dimming of a target star. The shape of this signal can be approximated as a box, i.e. the dimming initiates and terminates instantaneously with a constant flux during the dimming. BLS can also be adapted for microlensing studies by searching for \textit{non-repeating}, \textit{brightening} events. Though the magnification curve associated with microlensing of point-like sources is not well-approximated by a box, finite-source effects broaden this peak, resulting in light-curves that more closely resemble a box-like plateau (see \S\ref{sec:modeling} for an example displaying these effects, or Fig. 4 of \citet{Johnson_2022} for further discussion). Given that the majority of TESS targets lie within $\sim 3$ kpc of Earth, terrestrial-mass FFPs would produce events with significant finite-source effects, making the use of the BLS algorithm a reasonable initial means to flag high-SNR events of interest. BLS is also computationally inexpensive, which is necessary for searching the large TESS datasets. 

In particular, we used the GPU-optimized BLS algorithm implemented in \texttt{cuvarbase} \citep{cuvarbase} to search each light curve. For sensitivity to FFPs across a wide range of planetary masses, we search for events with durations as short as 17 minutes (five times the 200-second FFI cadence) and as long as 1 day. We set the period of the box signal to 30 days (longer than the total length of the sector) in order to force the algorithm to search for single events.

After BLS identified the duration ($\tau$) and central time ($t_{\mathrm{BLS}}$) of a possible signal, we estimated its signal-to-noise ratio (SNR) using
\begin{equation}\label{eqn:SNR}
    \mathrm{SNR} = \frac{\delta}{\sigma},
\end{equation}
\noindent where $\delta$ is the height of the event and $\sigma$ is the noise over the event duration. This is a rough estimate of SNR that could be improved by fitting a different shape such as a trapezoid \cite[e.g.,][]{Kipping}, which we will explore in future works to detect lower SNR events.

We estimated $\delta$ by finding the weighted mean of all data points over the event duration and subtracting the weighted mean of all data points at least one duration away from $t_{\mathrm{BLS}}$. We estimated $\sigma$ by taking into account correlated noise in the TESS light curves following the pink noise definition from \citet{Pont2006} as
\begin{equation}
    \sigma = \sqrt{\frac{\sigma_{w}^{2}}{n} + \sigma_{r}^{2}},
\end{equation}
where $\sigma_{w}$ is the white noise in the light curve, $\sigma_{r}$ is the red noise, and $n$ is the number of in-event data points. We take $\sigma_{w}$ to be equal to the weighted standard deviation of the light curve after masking out data within one event duration of $t_{\mathrm{BLS}}$. Following \citet{VARTOOLS}, we estimate $\sigma_{r}$ using the expression
\begin{equation}
    \sigma_{r}^{2} = \sigma_{\mathrm{bin}}^{2} - \sigma_{\mathrm{bin,exp}}^{2},
\end{equation}
\noindent where $\sigma_{\mathrm{bin}}$ is the weighted standard deviation of the residual light curve after binning in time with a bin-size equal to the duration of the event, and $\sigma_{\mathrm{bin,exp}}$ is the expected standard deviation of the binned light curve if the noise were uncorrelated in time. If $\sigma_{r}$ is estimated to be less than zero, we set $\sigma_{r} = 0$.

In order for a signal to be considered an initial candidate to be vetted further, we required $\mathrm{SNR} > 10$ and $n \geq 5$. Our requirement of $n \geq 5$ over the event duration differs slightly from other microlensing searches, which typically require a specific number of data points that are \textit{each} individually above some threshold, such as a departure $3\sigma$ from the baseline \citep{Johnson_2020}. Due to TESS’s rapid cadence, the number of datapoints associated with a given signal is often large, $\geq$ 20 for most potential FFP signals. By leveraging the large number of datapoints per event, signals with departures from baseline significantly below the per-datapoint uncertainty become detectable, as has been demonstrated in transit searches, which motivates our definition of detection criterion.

\subsection{Automated Flux-Level Vetting}

A total of 305,446 signals met our basic detection criteria and were sent through a first round of automated vetting, which involved computing a set of metrics to be compared to pass-fail thresholds. If a signal failed to meet any of our criteria, it was rejected as a possible microlensing candidate. We do not claim that these empirically-determined thresholds are optimal and, as is the case for any vetting procedure, they may have removed true signals as well as false positives. However, they successfully produced a dataset of potential signals amenable to manual inspection (\S\ref{sec:manual}).

\subsubsection{Initial Model Fitting}

\begin{figure*} \centering\includegraphics[width=0.45\linewidth]{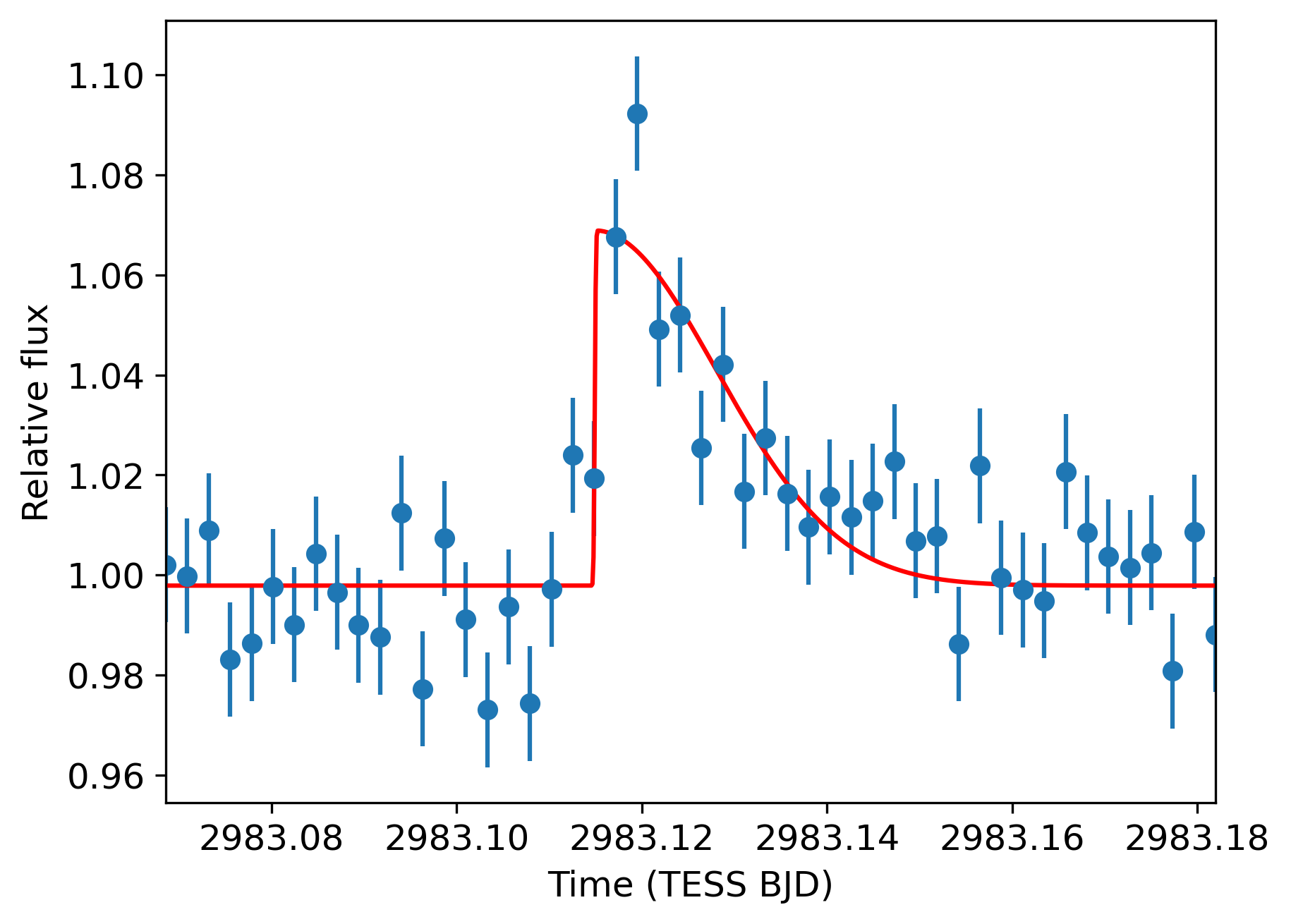}
\centering\includegraphics[width=0.43\linewidth]{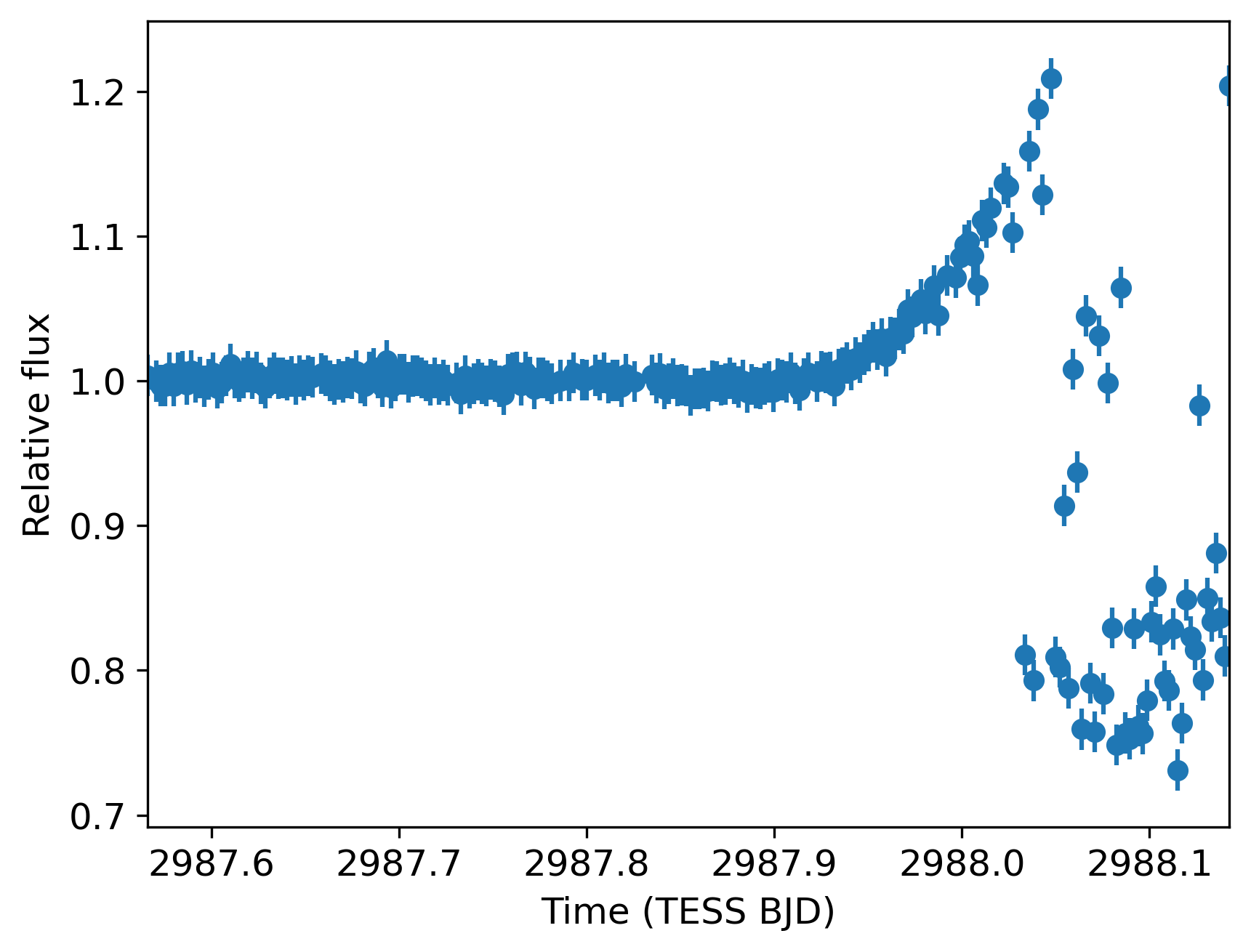}
    \caption{Examples of BLS-detected brightening events corresponding to flares (left) and scattered light (right). The flare event has a skew shape parameter $\alpha = 137$ based on a fit to a skew normal distribution (red line), which indicates strong asymmetry. The scattered light event reflects the most common type of microlensing false positive in TESS data, representing $\sim$80\% of BLS-detected events.} \label{fig:skew}
\end{figure*}

Each signal was first fit to three different models: a straight line parameterized by a slope and offset; a skew normal distribution parameterized by a central time, amplitude, scale ($\omega$), and shape ($\alpha$) (see left panel of Figure \ref{fig:skew}); and a simple point-source lens model parameterized by $t_{0}$, $u_{0}$, and $t_{E}$. Fitting was performed using the \texttt{lmfit} Python package \citep{LMFIT}. Data more than $2\tau$ from $t_{\mathrm{BLS}}$ were ignored in order to speed up the fitting process. 

An event was rejected if a point lens was a worse fit compared to a straight line according to the Bayesian Information Criterion (BIC) of each fit, i.e., $\mathrm{BIC}_{\mathrm{lens}} > \mathrm{BIC}_{\mathrm{line}}$. A total of 9,993 signals failed this criteria.

\subsubsection{Edge Effects and Gapped Events}

BLS occasionally triggers at the edge of data gaps, especially if such edges correspond to rapid increases in brightness due to scattered light (see right panel of Figure \ref{fig:skew}), a well-known systematic effect within TESS data.\footnote{\url{https://archive.stsci.edu/missions/tess/doc/TESS_Instrument_Handbook_v0.1.pdf}} A number of events also lack a significant amount of data either during the brightening, or just before and/or after. To identify these cases, we computed the fraction of the light curve near the event that actually contained data. We found $n_{\mathrm{exp}}$, the number of data points expected within $2 \tau$ of $t_{\mathrm{BLS}}$ given the 200-second cadence of the observations, as well as $n_{\mathrm{obs}}$, the number of data points actually observed over that time. We rejected 242,697 signals with $n_{\mathrm{obs}}/n_{\mathrm{exp}} < 0.9$. Given that this cut rejected 79\% of all vetted signals, edge effects and gapped events constituted the majority of TESS microlensing false positives.

\subsubsection{Non-Unique Events}
\label{sec:unique}

Microlensing events are intrinsically rare, arising from the chance alignment of source, lens, and observer. As such, the presence of multiple microlensing-like features at other times in a light curve suggest short-term periodic or quasi-periodic stellar variability,  low-SNR noise/instrumental systematics, or some other false positive. This provides a means of mitigating such false positives by searching for unique features in a given light-curve. To quantify uniqueness, we adopted elements from the Model-Shift Uniqueness Test developed for Kepler transiting exoplanet candidate vetting \citep{Coughlin2017}. In summary, we computed an SNR time series by finding the SNR of events centered at each point in the light curve over the duration of the BLS-detected signal. We measured the significance of the primary signal ($\mathrm{SNR}_{\mathrm{pri}}$) as the largest SNR value in the SNR time series within a half event duration of the BLS detection. The most significant event at least two event durations from the primary was labeled the secondary event ($\mathrm{SNR}_{\mathrm{sec}}$), and the next most significant event at least two event durations from both the primary and secondary was labeled the tertiary event ($\mathrm{SNR}_{\mathrm{ter}}$). Finally, the most significant negative event ($\mathrm{SNR}_{\mathrm{neg}}$), representing the largest flux decrease at least three event durations from the primary and secondary, was also labeled.

We rejected signals for which the primary event was not sufficiently unique, as quantified by $\mathrm{SNR}_{\mathrm{pri}} < 10$ (indicating that it is not unique compared to the noise) or  $\Delta\mathrm{SNR} < 2$ compared to the secondary, tertiary, or most negative events (indicating that it is not unique compared to other events at the same timescale). A total of 19,811 signals failed this criteria. 

The ability of such a test to rule out false positives is limited due to TESS's 27.4-day observation duration, as signals arising outside of this timespan are unobserved. This is in constrast to dedicated microlensing surveys, which rely on significantly longer baselines in order to establish the variable characteristics of source stars.

\subsubsection{Flares}\label{sec:flares}

Flares are typically highly asymmetric brightening events characterized by a rapid increase in flux followed by a gradual decay, in contrast to the symmetric, bell-shaped peaks of microlensing events. We used the results of the skew normal distribution fit to identify this class of false positives. We rejected 205,230 signals with a skew shape of $|\alpha| > 2$, suggesting strong asymmetry. Half of these also failed the edge effect/gapped event test, indicating that they were likely asymmetric events caused by scattered light. We also rejected 10,524 signals with a skew scale of $\omega < 0.01$ days, as these are consistent with extremely short-duration flares.

\subsection{Asteroid Rejection}

A total of 31,885 signals passed all previous tests, and the most common remaining false positive scenario was brightening events caused by passing asteroids. We employed three major tests to reject asteroids: \textit{(1)} cross-matching our candidates with known objects in the Solar System, \textit{(2)} analyzing each event at the pixel level, and \textit{(3)} searching for indications of moving bodies by matching correlated events across RA and Dec space.

\textit{(1) Known asteroids:} We queried the NASA/JPL Small-Body Identification API\footnote{\url{https://ssd-api.jpl.nasa.gov/doc/sb_ident.html}} for all known small bodies seen by TESS within the entire region in RA and Dec covered by Sectors 61--65 in each camera and CCD combination, extracting all objects brighter than 21 mag seen between 2962.79 and 3095.29 TESS BJD in steps of 2.5 days. We found 41,827 unique objects. Ephemerides for each were generated in 30-minute intervals using the JPL Horizons Solar System data and ephemeris computation service\footnote{\url{https://ssd.jpl.nasa.gov/horizons/}}. We then calculated the angular separation and difference in time between each asteroid and each surviving signal. We rejected 29,024 signals with angular separation of $<0.07^{\circ}$ and time difference of $< 0.2$ days.

\begin{figure}
    \includegraphics[width=\linewidth]{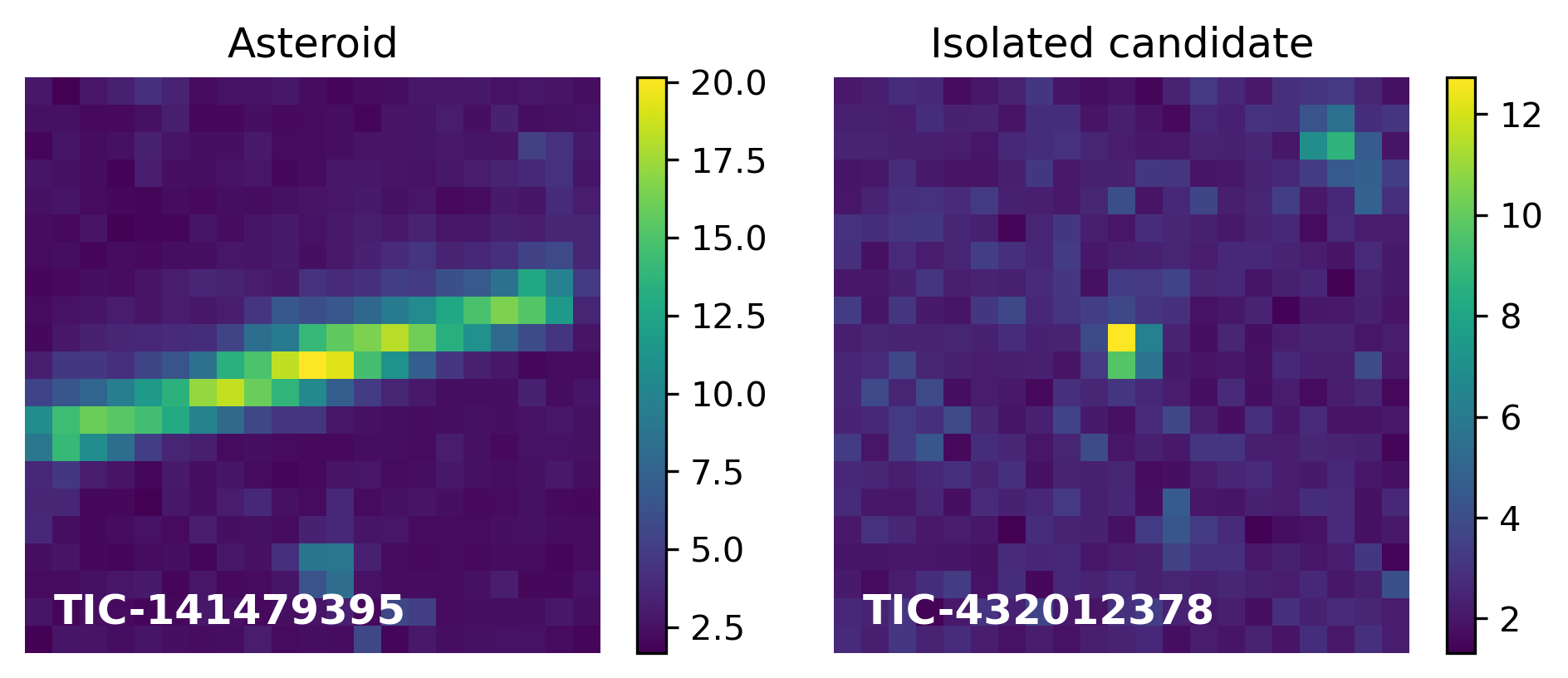}
    \caption{Pixel-level data demonstrating typical signatures for asteroids (left) compared to isolated events (right), using examples associated with TIC-141479395 and TIC-432012378, respectively. These data were produced by subtracting the mean out-of-event image from each image within $1.5\tau$ of the center of a given event, and finding the maximum value of each pixel.}\label{fig:asteroids}
\end{figure}

\textit{(2) Pixel-level analysis:} Hundreds of asteroids were not rejected by this automated cut due to incompleteness of the Small-Body Database. We further removed asteroids by inspecting each event at the pixel level. Using $21\times21$ pixel cutouts centered on each target star produced by \texttt{tesscut} \citep{tesscut}, we computed the mean of all out-of-event images to represent the typical field near a source star when no brightening or dimming events are occurring on average. Out-of-event times were defined as those between $0.5\tau$ and $1.5\tau$ from the center of each event. We then subtracted the mean out-of-event image from each individual image in order to find the pixels undergoing the largest brightening at each point in time. We searched for changes in the locations of the brightest pixels within $1.5\tau$ of the event, which are consistent with an asteroid moving across the TESS pixels. Examples showing the maximum values of each pixel, for both an asteroid and a likely isolated signal, are provided in Figure \ref{fig:asteroids}, highlighting the clear difference between the two scenarios.

We also produced difference images, employing the same techniques commonly used for transiting exoplanet vetting \citep{Bryson2013}, with the TESS difference imaging codebase \texttt{transit-diffImage}.\footnote{\url{https://github.com/stevepur/transit-diffImage}} Difference images are used to confirm that an event is co-located with the target star, rather than being due to a nearby contaminant. Nearby contaminants can include passing asteroids, other stars in neighboring pixels, or even stars in the same pixel as the target, given the large $21\times21^{\prime\prime}$ size of TESS pixels. Moving objects will appear smeared out in difference images, while on- and off-target microlensing events will remain isolated.

\textit{(3) RA/Dec event correlation:} As a final check, we inspected the RA-Dec distribution of our events to search for extended trails as the objects move across the TESS field and cause brightening events on different stars. We identified multiple trails, all of which also corresponded to asteroids identified through our pixel-level analysis. Between our pixel-level and RA-Dec investigations, we rejected another 2861 signals as likely asteroids. 

Note that we initially implemented the single-linkage clustering algorithm described by \citet{2021MNRAS.505.5584M} to automatically flag these correlated structures. However, we found that the low relative stellar density of the TESS field of view with respect to K2 Campaign 9 limited the efficacy of this method, and we removed it from our vetting procedure. In future work to further automate the process, we plan to instead incorporate shift-and-stack methods \cite[e.g.,][]{Holman2019,Payne2019}, which have already been used to detect asteroids with TESS FFI data \citep{Woods2021}.

\subsection{Manual Inspection}\label{sec:manual}

We manually inspected both light curves and difference images for the surviving 452 candidates in order to confirm that the events appeared symmetric, unique, and consistent with our expectations for on-target lensing events. Most events that quickly failed visual inspection were obvious detrending artifacts caused by discontinuities in the QLP light curves, or were caused by stellar variability that fell beyond the thresholds of the cuts used during the automated vetting procedure. In some cases, inspection of the difference images revealed that the signal was offset from the target star. Such events were further investigated using the light curve of the correct source; all turned out to be variable stars.

Based on flux- and pixel-level data, we found two high-SNR signals associated with TIC-123147666 and TIC-107150013 that had light-curve morphologies consistent with expectation of a short-duration microlensing event in Sector 61. However, a single TESS sector is insufficient for identifying repeating events that occur on timescales near or longer than 27 days, such as periodic brightening events caused by tidal distortion in highly eccentric binary star systems \cite[e.g., KOI-54;][]{KOI54}. 

\begin{figure*}
    \centering\includegraphics[width=0.9\linewidth]{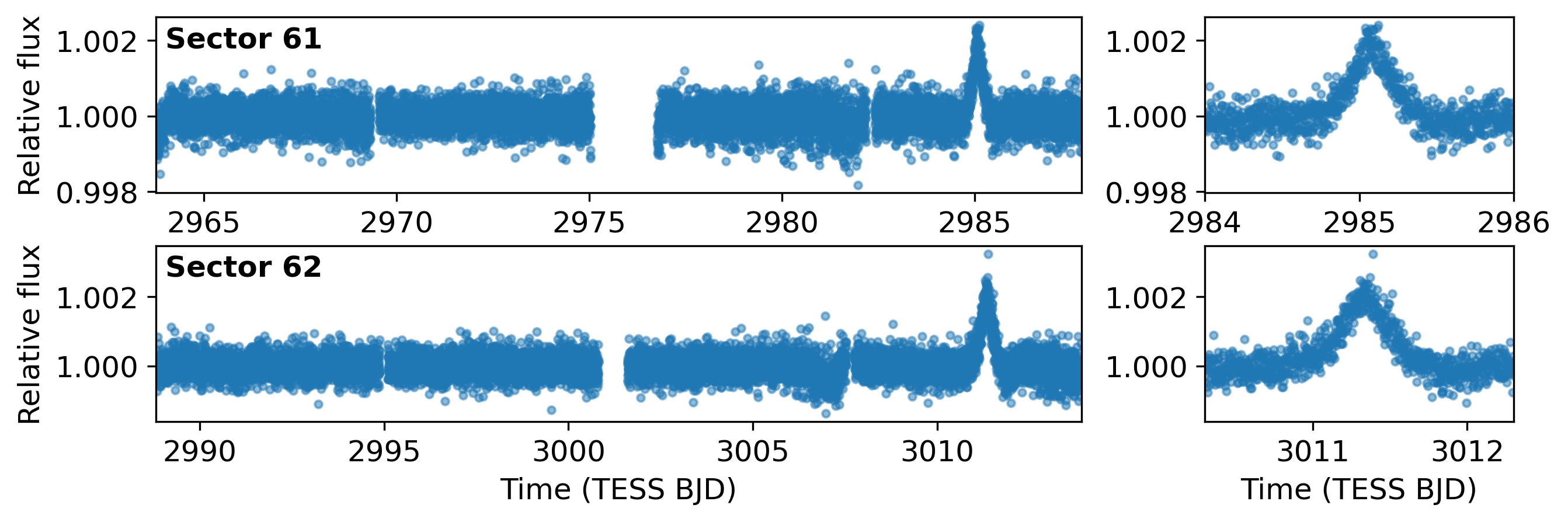}
    \caption{Detrended QLP light curves from Sector 61 (top) and Sector 62 (bottom) for TIC-123147666, which feature a brightening event that was initially identified in Sector 61 as a potential microlensing event before repeating events were identified in other TESS sectors. This event is likely caused by tidal distortion in a highly eccentric binary star system. The heartbeat nature of the event is more obvious in Sector 62, further emphasizing the need for multiple sectors when performing microlensing searches with TESS.}\label{fig:123147666}
\end{figure*}

We visually inspected all available TESS observations for both of these signals in search of repeated events, and found that TIC-123147666 (Figure \ref{fig:123147666}) featured similar brightening events in all other TESS sectors (Sectors 7, 8, 34, 35, 62, 87, and 88), consistent with a period of 26.2 days. This star was also previously identified as having a proper motion anomaly between the long-term proper motion vector and Gaia DR2 and HIPPARCOS measurements, indicative of the presence of a perturbing secondary object \citep{PMa}. TIC-123147666 also has a high Renormalized Unit Weight Error (RUWE) value (8.10) according to Gaia DR3 observations \citep{Gaia,GaiaDR3}, consistent with having a companion \citep{Lindegren2018,Lindegren2021}. We therefore classify TIC-123147666 as a false positive likely due to tidal interactions with a binary companion. Upon revisiting several of our other high-SNR, non-asteroid signals that had been manually rejected due to lack of consistency with lens models, we found that many were associated with repeating events throughout other TESS sectors. Future vetting efforts could be substantially improved and further automated by incorporating multi-sector data earlier in the process.

\begin{figure}
    \centering
   \includegraphics[width=\linewidth]{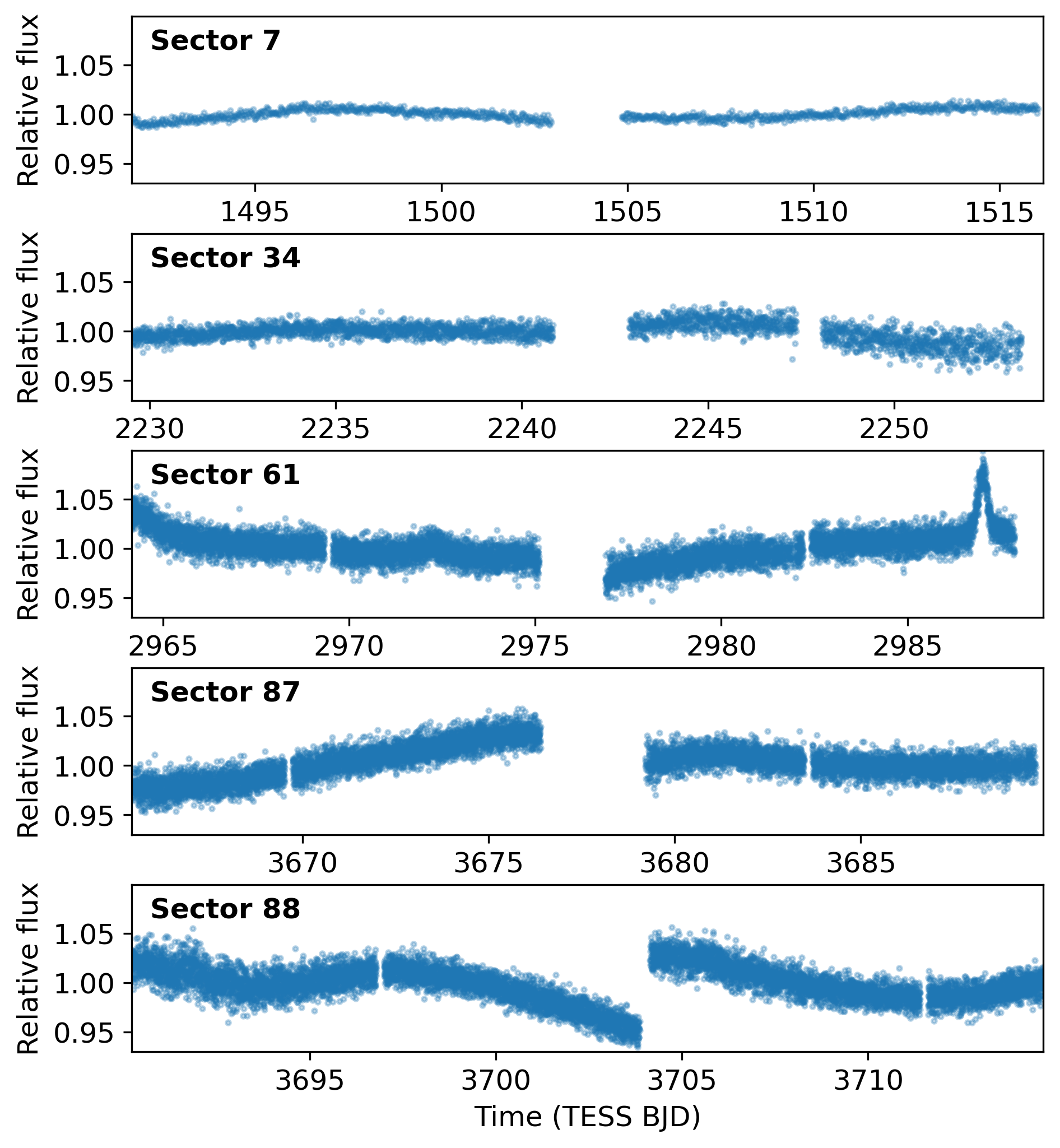}
   \caption{Pre-detrended QLP light curves from Sectors 7, 34, 61, 87, and 88 for TIC-107150013. There are no events with comparable significance to the event in Sector 61 at $\sim2987$ days. The increase in noise in later sectors is due to their shorter cadences, which are associated with higher per-point uncertainties. Sector 7 was observed at 30-min cadence, Sector 34 was observed at 10-min, and Sectors 61--88 were observed at 200-sec. The light curves potentially indicate stellar variability over timescales on the order of tens of days, but TESS alone is not able to constrain the amplitude and period of this variability due to systematic differences between TESS orbits.}\label{fig:multisectors}
\end{figure}

In contrast to TIC-123147666, we found no repeating events in other sectors for TIC-107150013 (Sectors 7, 34, 87, and 88). Furthermore, its low RUWE value (0.99) is consistent with a single star, disfavoring the presence of an unresolved binary companion associated with this source. There is potentially stellar variability at a timescale of order tens of days visible in the TESS data (Figure \ref{fig:multisectors}), though TESS is not well-suited to analysis of periodic variability longer than $\sim13$ days due to systematics introduced by the telescope orbit \citep[e.g.][]{Hedges2020, Claytor2022}. However, comparison to longer-baseline observations made by the long-running Optical Gravitational Lensing Experiment (OGLE) indicate that there is some degree of $\sim40$-day periodic stellar variability. \citet{Mroz2024} suggests that this may be due to the presence of stellar spots, indicating potential magnetic activity on this star. Additionally, \citet{Mroz2024} presents two flares from the catalog compiled in \citet{Iwanek2019} that were qualitatively identified as being symmetric, and uses these events to make a rough estimate of the symmetric flare rate on giants in TESS data. However, as we show in \S\ref{sec:results}, a Bayesian evaluation of which model is a better fit (between microlensing and flares) provides a quantitative means of more robustly addressing flare contamination in our data. We perform this analysis on the TIC-107150013 event as well as a dataset of flares with TESS's fast cadence and high photometric precision to assess the level of contamination of symmetric flares in our microlensing search.

\section{Results and Discussion}\label{sec:results}

\begin{figure*}
\centering
    \includegraphics[width=\textwidth]{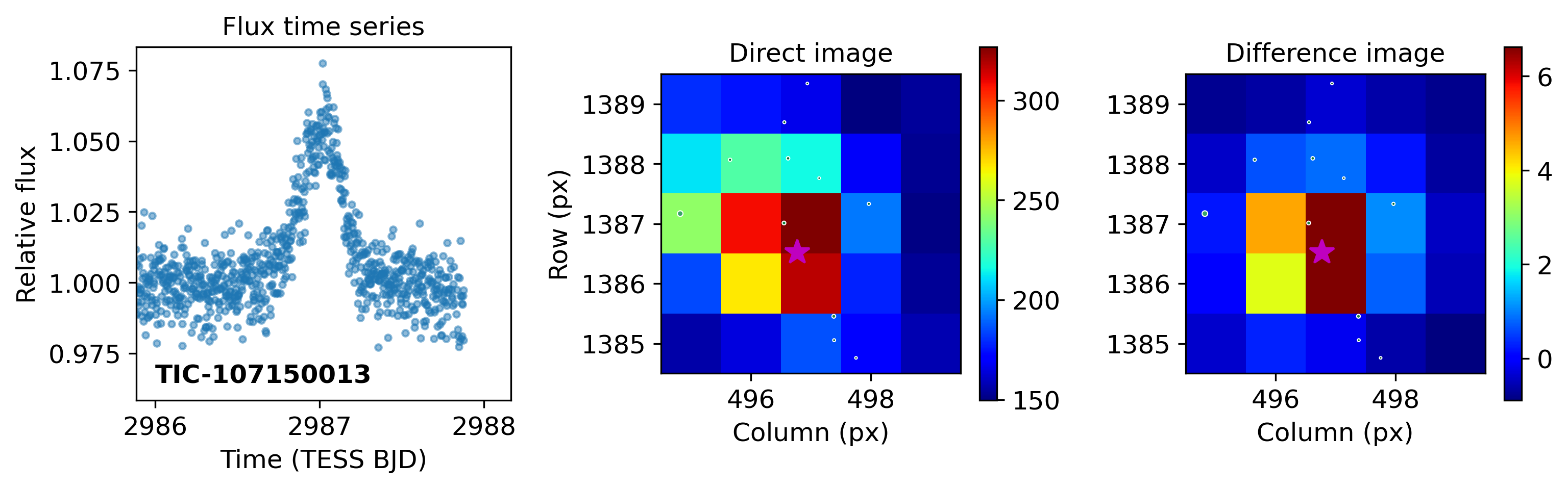}
    \caption{Flux- and pixel-level data associated with the source star TIC-107150013, which features a high-SNR ($\mathrm{SNR} = 31$) transient signal at 2987 days. The direct image reflects the typical field near the target (pink star), while the difference image reflects the pixels undergoing the largest brightening over the event. White circles show the locations of all stars down to 4 magnitudes fainter than the target star, with sizes scaled by brightness. The difference image is consistent with the signal being co-located with the target star.}\label{fig:cands}
\end{figure*}

TIC-107150013 is located at a distance from Earth of $d_{s} = 3193\pm153$ pc based on parallax from Gaia DR3 (Table \ref{tab:star}). Lacking a stellar radius from Gaia DR3, we adopt $R_{s} = 12.91_{-1.51}^{+0.96}~R_{\odot}$ from Gaia DR2 \citep{Gaia,DR2}. The star lies $5.03^{\circ}$ below the Galactic plane, making it a relatively crowded region of the sky that is not very well explored by microlensing surveys; however, at $T = 13.10$ mag, TIC-107150013 is the brightest star within $1^{\prime}$. The observed short-duration brightening event occurs at $t_{\mathrm{BLS}} = 2987.03$ days, was detected by BLS with a duration of $\tau = 0.38$ days, and has a high SNR of 31 (Figure \ref{fig:cands}). In order to further assess the event, we performed both microlensing fits and fits of stellar flare models to the data. We analyzed the same signal using multiple light-curve extraction methods and, in the case of the microlensing fit, performed dedicated lens modeling including second-order effects arising from the non-negligible angular extent of the source and limb-darkening.

\begin{table*}
    \centering
    \caption{Stellar Parameters for TIC-107150013}
    \begin{tabular}{l|l|l|l}
        \hline\hline
        Parameter & Value & Description & Source \\ 
        \hline\hline
        \multicolumn{4}{c}{\textit{TIC Parameters}} \\
        \hline
        ID & 107150013 & TESS Input Catalog ID & TICv8.2 \\
        $T_{\rm eff}$ & $4115\pm123$ & Effective temperature (K) & TICv8.2\\
        $R_{s}$ & $12.91_{-1.51}^{+0.96}$ & Stellar radius ($R_{\odot}$) & Gaia DR2 \\
        \hline
        \multicolumn{4}{c}{\textit{Astrometric Parameters}} \\
        \hline
        RA & 07:22:30.806 & Right ascension (J2016) & Gaia DR3 \\
        Dec & -25:29:15.77 & Declination (J2016) & Gaia DR3 \\
        $\bar{\omega}$ & $0.313\pm0.015$ & Parallax (mas) & Gaia DR3\ \\
        $\mu_{\alpha}$ & $-0.395\pm0.010$& Proper motion right ascension (mas~yr$^{-1}$) & Gaia DR3 \\
        $\mu_{\delta}$ & $0.260\pm0.016$ & Proper motion declination (mas~yr$^{-1}$) & Gaia DR3 \\
        \hline
        \multicolumn{4}{c}{\textit{Photometric Parameters}} \\
        \hline
        $T$ & $13.0985\pm0.0071$ & TESS band magnitude (mag) & TICv8.2 \\
        $B$ & $17.228\pm0.162$ & $B$ band magnitude (mag) & UCAC4 \citep{UCAC4}\\
        $V$ & $14.922\pm0.103$ & $V$ band magnitude (mag) & UCAC4\\
        $G$ & $14.017\pm0.002$ & Gaia $G$ band magnitude (mag) & Gaia DR3 \\
        $J$ & $11.63\pm0.026$ & $J$ band magnitude (mag) & 2MASS \citep{2MASS} \\
        $H$ & $10.91\pm0.022$ & $H$ band magnitude (mag) & 2MASS \\
        $K_{s}$ & $10.64\pm0.023$ & $K$ band magnitude (mag) & 2MASS \\
        $W1$ & $10.453\pm0.022$ & $W1$ band magnitude (mag) & WISE \citep{WISE}\\
        $W2$ & $10.538\pm0.020$ & $W2$ band magnitude (mag) & WISE \\
        $W3$ & $10.373\pm0.075$ & $W3$ band magnitude (mag) & WISE \\
        \hline
\end{tabular}\label{tab:star}
\end{table*}

\subsection{Alternative Light Curve Extractions}\label{sec:lcs}

\subsubsection{TESS-Gaia Light Curves}

QLP produces light curves using an aperture photometry approach combined with difference imaging. A reference image is constructed using the median combination of good quality images and this reference is subtracted from each frame to produce difference images. Aperture photometry is performed on the difference images to produce a difference flux time series, and the TESS magnitude provided in the TIC is used to scale the difference fluxes relative to the star's average flux. The end result is a de-blended flux time series assuming any variations observed are associated with the target star \citep{QLP1}. This approach relies on an accurate estimate of the star's TESS magnitude, and the amplitude of a microlensing event will be inaccurate if the target star is not the true source.

To test the sensitivity of the interpretation of the event to an alternative method of data reduction, we extracted a new Sector 61 light curve using the TESS-Gaia Light Curve (TGLC) procedure \citep{TGLC} with a $50\times50$ pixel FFI cutout size. TGLC produces both aperture and point spread function (PSF) high-precision light curves corrected for contamination from nearby stars based on star positions and magnitudes from Gaia DR3. Here we adopt the TGLC aperture light curves, which offer more reliable amplitude estimates for dim stars in crowded fields like TIC-107150013 \citep{TGLC}. The TGLC aperture light curves are produced by performing aperture photometry on a $3\times3$ pixel aperture centred on the target star. To correct for contamination, TGLC estimates the median total flux of nearby stars in the Gaia DR3 catalog as well as the percentage of the target star's light that falls on the aperture based on the shape of an effective PSF model. The aperture light curve’s median is shifted to the Gaia-predicted median multiplied by this percentage, resulting in a de-blended light curve corrected for the contamination of all known stars resolved by Gaia DR3.

\begin{figure}
    \centering
    \includegraphics[width=\linewidth]{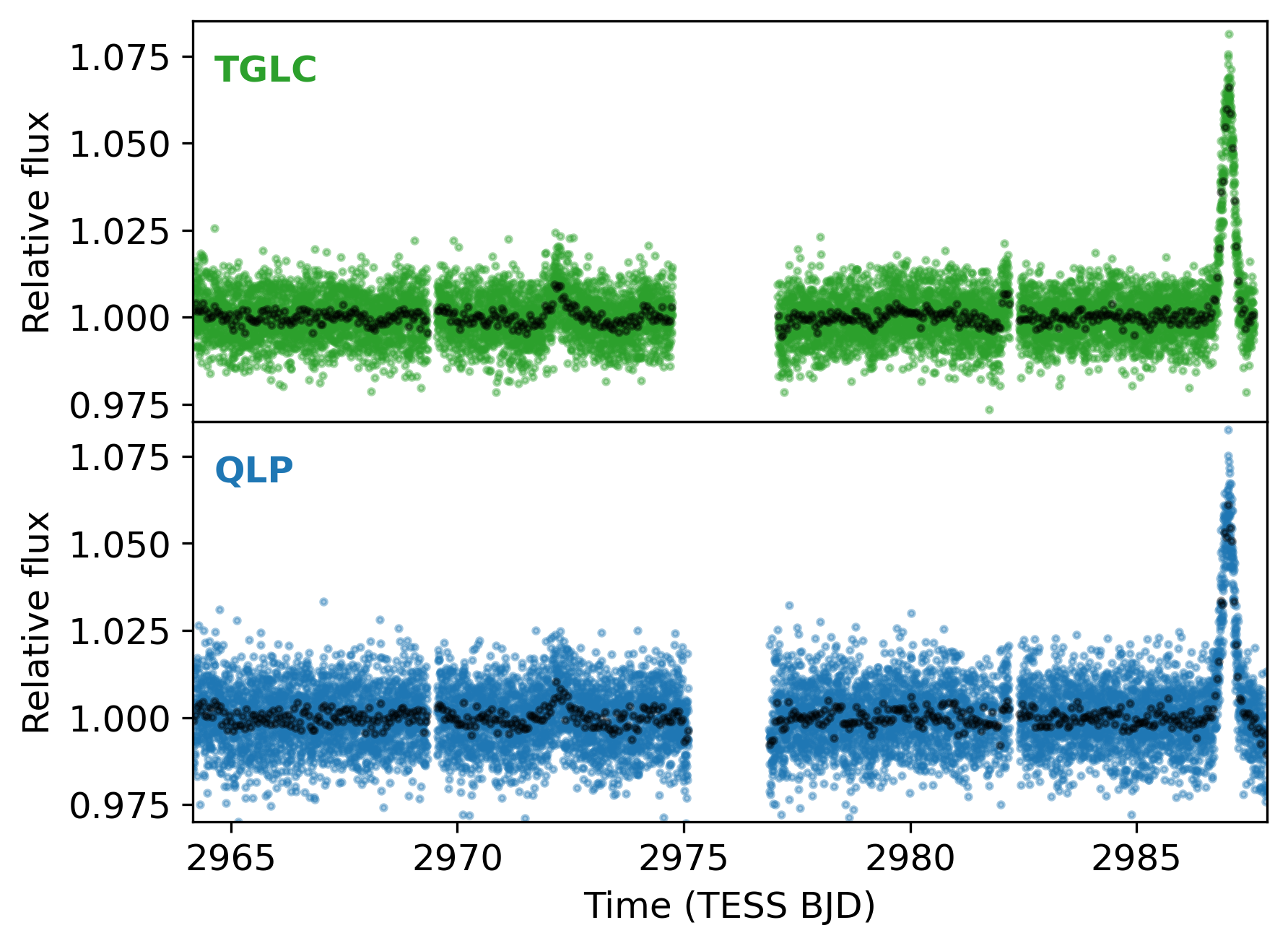}
    \caption{Detrended Sector 61 light curves for TIC-107150013, as extracted by TGLC (top) and QLP (bottom). The black points show the data binned in 1-hr bins. The signal at 2987 days is visually consistent between the two light curves, with similar shape, duration, and amplitude.}\label{fig:tglc_qlp_det}
\end{figure}

Because our original detrending process can be destructive to the shapes of high-SNR lensing events, we also re-detrended each light curve while masking out data within $0.5\tau$ of the event to minimize distortion of the microlensing signal. The re-detrended QLP and TGLC light curves are shown in Figure \ref{fig:tglc_qlp_det}, with the signal at 2987 days clearly visible in both datasets. Because the TGLC light curve has better precision, with an out-of-event flux standard deviation of 6448 ppm compared to 8603 ppm in the QLP light curve, we adopt the TGLC light curve for our preferred results in further analysis.

The re-detrended light curves revealed the presence of an additional small transient feature at $t = 2972$ days that was below the threshold for significant detection in the original QLP extraction. As discussed in \S\ref{sec:unique}, microlensing events are expected to be isolated. Hence, the presence of this feature suggests some form of false positive contamination. In \S\ref{sec:modeling}, we first operate under the assumption that the primary event is due to microlensing to highlight some of its more unusual aspects, then in \S\ref{sec:falsepositive} discuss this interpretation in the context of false positive contamination.

\subsubsection{SPOC Light Curves}

We also obtained light curves for TIC-107150013 that were processed by the Science Processing Operations Center \citep[SPOC;][]{SPOC} at the NASA Ames Research Center. SPOC produces light curves at shorter cadences than the FFIs for a pre-selected list of $\sim20,000$ targets each sector. TIC-107150013 was not selected for SPOC processing in Sectors 7, 34, and 61, but we obtained 20-sec cadence observations in Sectors 87 and 88 through a Director’s Discretionary Targets (DDT) Proposal (DDT079, PI: Kunimoto). While these observations are long after the Sector 61 event, the 20-sec cadence observations are valuable for characterizing any potential flaring nature of the host star, where flares are potential microlensing false positives. 

\begin{figure*}
\centering
    \includegraphics[width=0.9\textwidth]{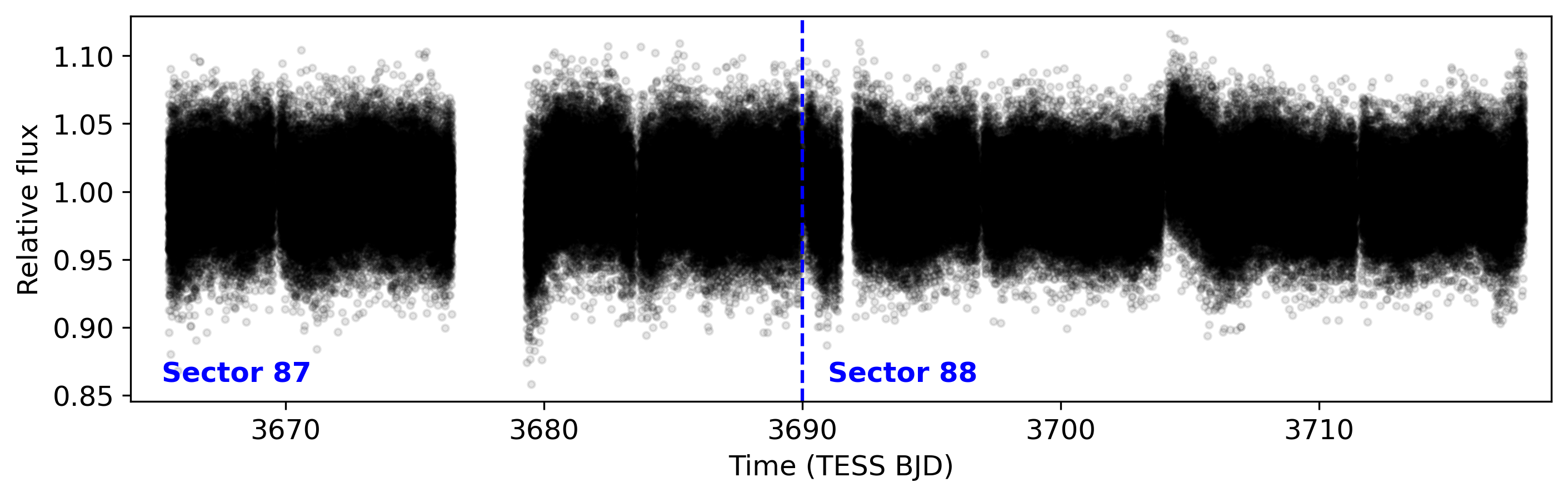}
    \caption{20-sec cadence SPOC PDCSAP light curve for TIC-107150013 from Sectors 87 and 88. There are no repeated signals resembling the event from Sector 61, nor any indications of flares.}\label{fig:spoc}
\end{figure*}

In Figure \ref{fig:spoc}, we show the 20-sec cadence SPOC Pre-search Data Conditioning Simple Aperture Photometry \cite[PDCSAP;][]{Smith2012,Stumpe2012,Stumpe2014} light curves for Sectors 87 and 88, with all data points with a nonzero quality flag removed. Differences in the appearance of long-term variability compared to the QLP light curves (Figure \ref{fig:multisectors}) can be attributed to differences between light curve extraction procedures and systematics corrections; however, there are no repeated signals resembling the event from Sector 61, nor any indications of flares.

\subsection{Microlensing Interpretation}
\label{sec:modeling}

In this section, we treat the signal as a microlensing event. We fit the signal to a microlensing template and use the results to check the consistency of this event in three ways: (1) consistency with the expectation for a light curve produced by a low-mass FFP, (2) consistency of the implied mass and velocity with expectations for nearby FFPs, and (3) consistency of the implied event rate with the expected abundance of FFPs. 

\subsubsection{Consistency with Light Curve Morphology}\label{sec:lensingmodels}

We fit both QLP and TGLC light curves using point-source point-lens (PSPL) and finite-source point-lens (FSPL) models with linear limb-darkening, as implemented by the \texttt{pyLIMA} lensing model package \citep{pylima}. For the FSPL fits we used the \texttt{pyLIMA} \texttt{FSPLargemodel}, which is appropriate for the finite-source magnification of large stars. We adopted a linear limb-darkening coefficient of 0.68 based on TESS values estimated by \citet{Claret2017} for relatively cool giant stars ($\log g \sim 2.5$, assumed given the star's giant nature; $T_{\mathrm{eff}} \sim 4100$ K, based on the star's $T_{\mathrm{eff}}$ provided in the TICv8.2). We also fit for the flux contributions from both the source and possible blended stars ($f_{\mathrm{source}}$ and $f_{\mathrm{blend}}$, respectively). Parallax effects due to the motion of the satellite are negligible on these timescales for all realistic lens masses and distances  and thus were not included in the fit. 

Ultimately, we performed five-parameter PSPL fits parameterized by $t_{0}$, $u_{0}$, $t_{E}$, $f_{\mathrm{source}}$, and $f_{\mathrm{blend}}$, and six-parameter FSPL fits with the additional parameter $\rho$. We required $f_{\mathrm{source}} \geq 0$ and assigned a Gaussian prior to $f_{\mathrm{blend}} \sim \mathcal{N}(0,\sigma^{2})$ where $\sigma$ is 10 times the photometric uncertainty ($\sigma = 55~e^{-}s^{-1}$ and $74~e^{-}s^{-1}$ for TGLC and QLP, respectively). The prior is centered on 0 based on our expectation that the blending fraction should be low, given that (a) the light curves produced by the TGLC and QLP extractions have already undergone de-blending of resolved stars (\S\ref{sec:lcs}), (b) TIC-107150013 is very bright in comparison to unresolved background stars, dominating the flux, and (c) TIC-107150013 is consistent with being a single star based on Gaia DR3 astrometric data (RUWE = 0.99). We adopted uniform priors for all other fit parameters, as shown in Table \ref{tab:results}.

To explore the parameter space, we used the \texttt{emcee} ensemble sampler \citep{ForemanMackey2013,Goodman2010} with 50 walkers. Following \citet{Goodman2010}, we compute the integrated autocorrelation time for each fit parameter in order to assess convergence. We consider the chains converged if the number of steps is at least 50 times longer than the average autocorrelation time, and throw out burn-in steps equal to twice the maximum autocorrelation time. The TGLC light curve fits converged after 2700 and 3400 steps for PSPL and FSPL models, respectively, while the QLP light curve fits converged after 2800 steps and 3900 steps.

\begin{figure*}
    \centering
    \includegraphics[width=0.8\linewidth]{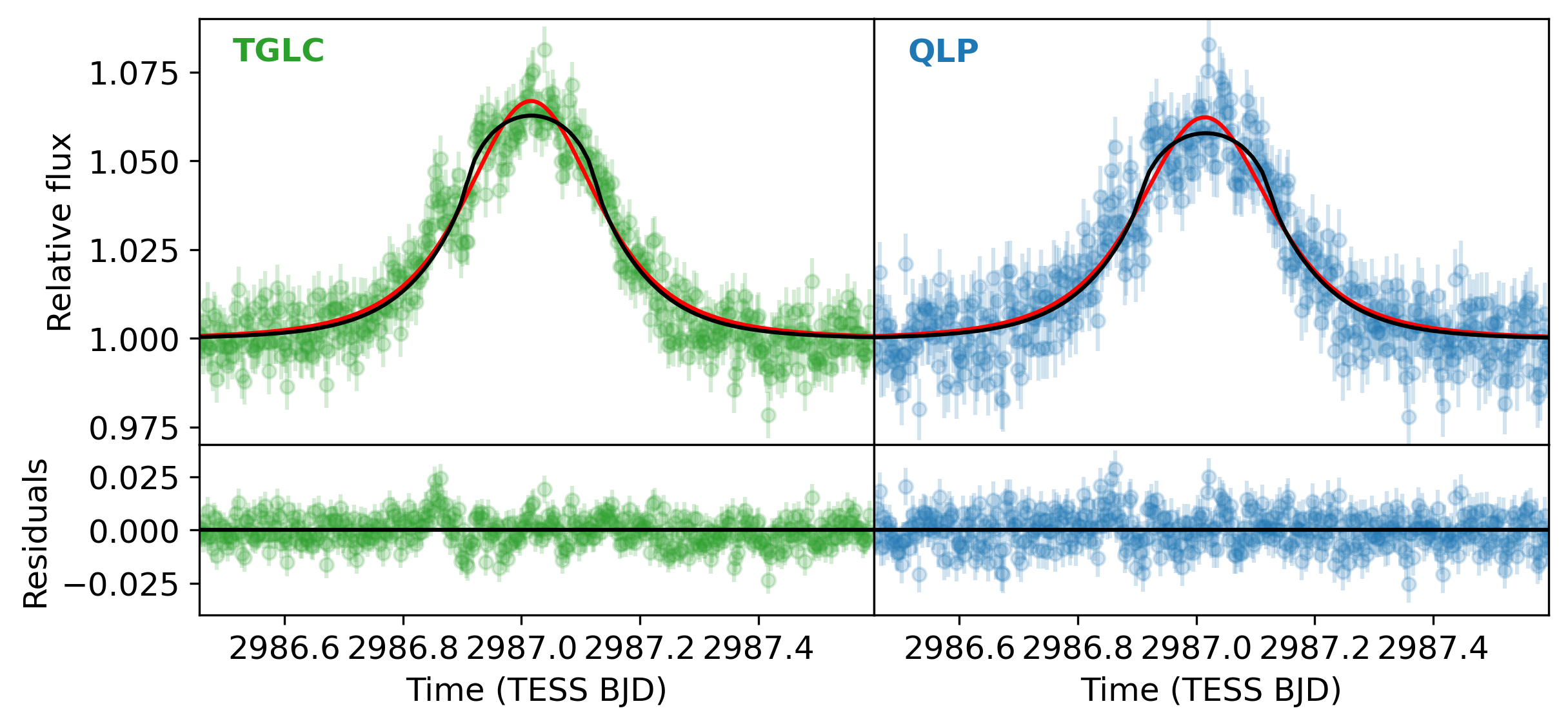}
    \caption{The TESS event associated with the source star TIC-107150013, as seen in TGLC (left) and QLP (right) sector 61 light curves. The red lines show the median PSPL fits, while the black lines show the median FSPL model fits. Based on comparing BIC values for each result, the FSPL fits are overwhelmingly preferred. The residuals show the light curve after removal of the FSPL model.}\label{fig:models}
\end{figure*}

\begin{figure*}
    \centering
    \includegraphics[width=\linewidth]{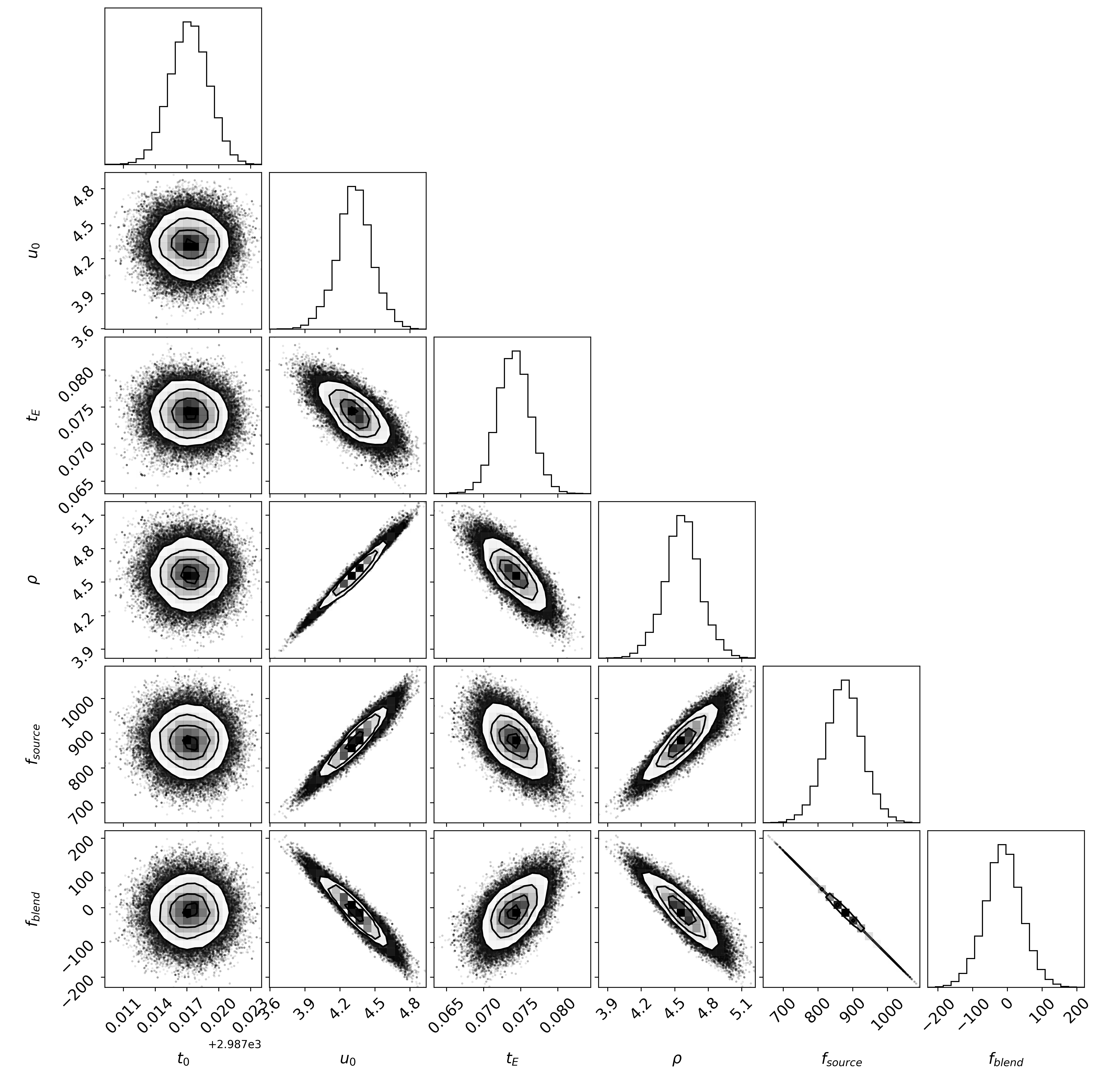}
    \caption{Corner plot displaying the \texttt{emcee} posteriors from the FSPL fit to the TGLC light curve of TIC-107150013.}\label{fig:corner}
\end{figure*}

Our fit results are shown in Table \ref{tab:results} and Figure \ref{fig:models}. We include the BIC values for each model in Table \ref{tab:results} and use this as our basis for model comparison. We chose BIC over alternative metrics such as the Akaike Information Criterion (AIC) because BIC is well-suited to comparing the explanatory capabilities of competing models \citep{Shmueli2010}. BIC imposes a heavier penalty than AIC for model complexity, which helps guard against overfitting. Despite the increased number of fit parameters required by the FSPL model, we find that PSPL fits had significantly larger BIC values ($\mathrm{BIC}_{\mathrm{PSPL}} - \mathrm{BIC}_{\mathrm{FSPL}} = $ $21.0$ and $12.0$ for TGLC and QLP, respectively), indicating that the FSPL fits are overwhelmingly preferred \cite[$\Delta\mathrm{BIC} > 10$;][]{Kass01061995}. The TGLC corner plots for the FSPL fit are shown in Figure \ref{fig:corner}. Both fits give blended flux contributions consistent with 0, and recover statistically significant finite source effects, with $\rho = 4.57_{-0.15}^{+0.15}$ from TGLC and $\rho = 4.75_{-0.23}^{+0.23}$ from QLP.

\begin{table*}
    \centering
    \caption{Best-fit PSPL and FSPL model parameters based on the TGLC and QLP Sector 61 light curves of TIC-107150013, including $\theta_{E}$ derived from the FSPL fit results and stellar angular size. The central values are the median of the posteriors, with lower and upper uncertainties from the 16th and 84th percentiles. We also report the BIC values for each model, indicating that the FSPL fits are preferred.}\label{tab:results}
    \begin{tabular}{llcccc}
        \hline\hline
        Parameter & Prior & TGLC Fit (PSPL) & QLP Fit (PSPL) & TGLC Fit (FSPL) & QLP Fit (FSPL) \\ 
        \hline\hline
        $t_{0}$ (BJD - 2457000) & $\mathcal{U}(2986.6, 2987.4)$ & $2987.017_{-0.002}^{+0.002}$ & $2987.015_{-0.003}^{+0.003}$ & $2987.017_{-0.002}^{+0.002}$ & $2987.015_{-0.003}^{+0.003}$ \\
        $u_{0}$ & $\mathcal{U}(0, 10)$ & $1.93_{-0.04}^{+0.04}$ & $1.98_{-0.06}^{+0.06}$ & $4.33_{-0.15}^{+0.15}$ & $4.51_{-0.22}^{+0.21}$\\
        $t_{E}$ (days) & $\mathcal{U}(0.01, 1)$ & $0.088_{-0.002}^{+0.002}$ & $0.088_{-0.003}^{+0.003}$ & $0.074_{-0.002}^{+0.002}$ & $0.074_{-0.003}^{+0.003}$ \\ 
        $\rho$ & $\mathcal{U}(5\times10^{-5}, 10)$ & - & - & $4.57_{-0.15}^{+0.15}$ & $4.75_{-0.23}^{+0.23}$ \\
        $f_{\mathrm{source}}$ ($e^{-}s^{-1}$) & $\geq 0$ & $872_{-54}^{+54}$ & $867_{-72}^{+73}$ & $875_{-52}^{+52}$ & $875_{-74}^{+72}$\\
        $f_{\mathrm{blend}}$ ($e^{-}s^{-1}$) & $\mathcal{N}(0, \sigma^{2})$\footnote{$\sigma$ is equal to 10 times the photometric uncertainty ($\sigma = 55~e^{-}s^{-1}$ and $74~e^{-}s^{-1}$ for TGLC and QLP, respectively)} & $-8_{-55}^{+54}$ & $-3_{-73}^{+72}$ & $-11_{-52}^{+52}$ & $-11_{-72}^{+74}$\\
        \hline
        $\theta_{E}$ ($\mu\mathrm{as}$) & - & - & - & $4.07_{-0.48}^{+0.43}$ & $3.91_{-0.48}^{+0.45}$\\
        \hline
        BIC & & $56253.9$ & $62318.7$ & $56232.9$ & $62306.7$ \\
\end{tabular}
\end{table*}

There remains a small feature in the residuals around $t=2986.85$ days that is particularly visible in the TGLC light curve for the best-fit FSPL model. This may be due to systematic noise or a genuine brightening of the source. Beyond an FSPL model alone, we attempt to explain the event as a double flare in \S\ref{sec:falsepositive} and find that the FSPL interpretation is still strongly preferred. Hence, from these analyses, we conclude that the primary brightening event associated with this light curve is consistent with the expectations of a low-mass FFP microlensing event, making it impossible to eliminate as a false positive based on the local light curve morphology alone.

\subsubsection{Consistency with Physical Parameters}
\label{subsec:physical}

Though the reconstruction of finite source features is challenging owing to degeneracies with other fit parameters \citep{2022ApJ...927...63J}, the appearance of putative finite-source features in the light curve allows a direct estimate the Einstein angle that would be associated with the best-fit parameters, which in turn provides an estimate of the lens mass and relative velocity. We compare these to typical planetary masses and the local Galactic velocity distribution \citep{2018A&A...616A..11G}, as false positives would likely produce physical parameter estimates inconsistent with the expectations for a population of FFPs. Given TIC-107150013's distance of $d_{s} = 3193\pm153$ pc and physical radius $R_{s} = 12.91_{-1.51}^{+0.96}~R_{\odot}$,  we estimate its angular size as $\theta_{s} = R_{s}/d_{s}~\mathrm{rad} = 18.8_{-2.1}^{+1.9}~\mu\mathrm{as}$. Assuming $\rho = 4.57_{-0.15}^{+0.15}$ from the TGLC fit, this corresponds to an Einstein angle of $\theta_{E} = 4.07_{-0.48}^{+0.43}~\mu\mathrm{as}$. 

\begin{figure}
    \centering
    \includegraphics[width=\linewidth]{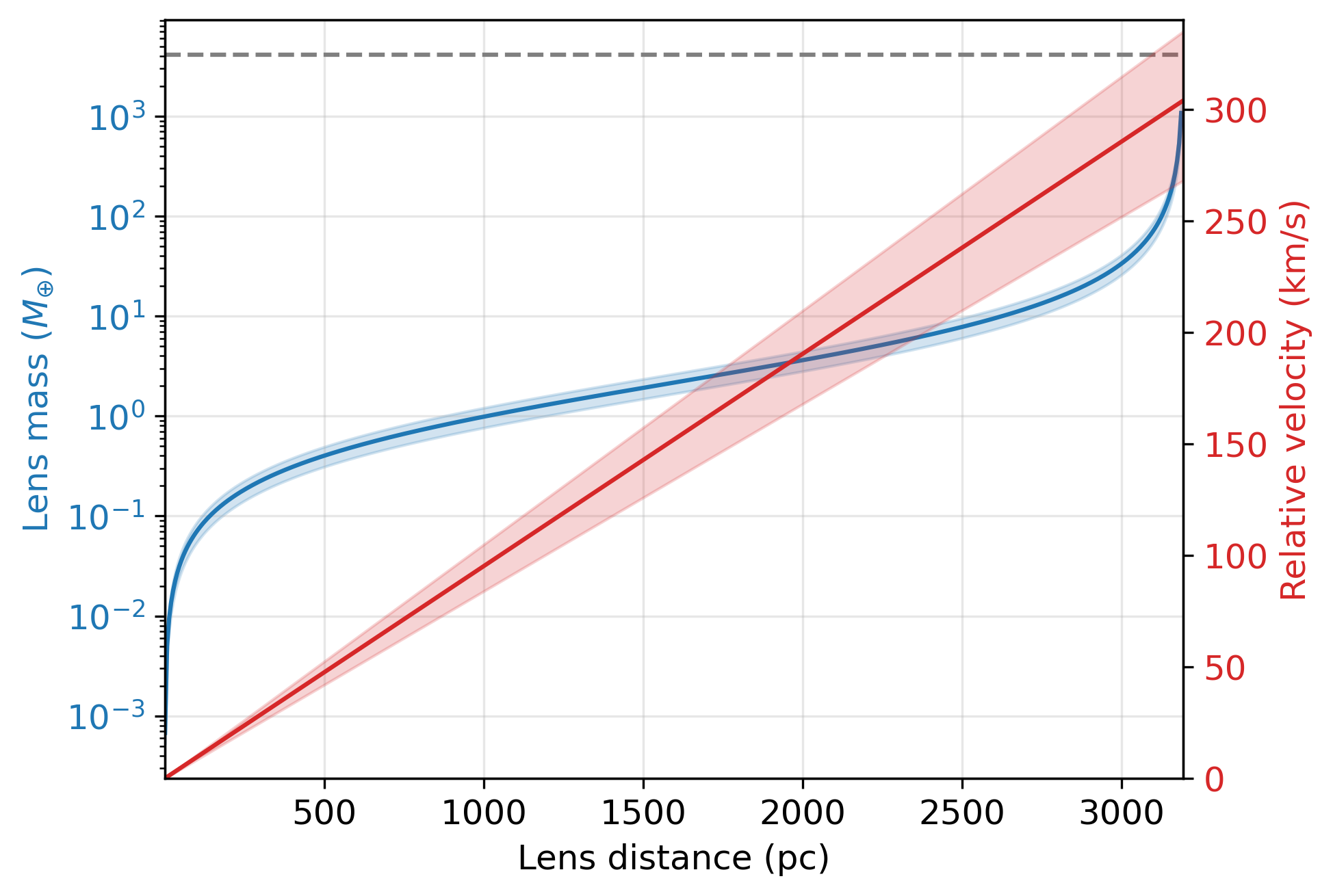}
    \caption{Hypothetical lens masses (blue) and relative velocities (red) at different lens distances based on the FSPL fits to the TGLC light curve. The shaded regions correspond to the 16th to 84th percentiles of the fit results for $\theta_E$ and $t_E$, with uncertainties from $R_{s}$ and $d_{s}$ propagated, while the solid line corresponds to the median values. Note that some regions are more likely than others, as discussed in \ref{subsec:physical}. The dotted grey line corresponds to 13 $M_{\mathrm{Jup}}$, which we assume is the largest possible planetary mass.}\label{fig:masses}
\end{figure}

In the following, we again emphasize that the calculations are carried out under the working assumption of a detected event for the sake of clarity. For an event with measured $t_E$ and $\theta_E$, there remains an inherent degeneracy between the physical parameters $M, v_T$, and $d_L$. As such, in Fig. \ref{fig:masses}, we plot both the mass (blue) and transverse velocity (red) as a function of lens distance. Without additional information about the system, e.g. a measurement of $d_L$ from parallax, this degeneracy cannot be fully broken to make individual estimates of each parameter. However, one can take a Bayesian approach and build a posterior for the lens mass and distance by adopting a prior over the distribution of FFPs, accounting for the likelihood of recovering the observed values of $\theta_E$ and $t_E$. We take the reasonable prior on $d_L$ that the FFP abundance tracks the stellar abundance along the line of sight to TIC-107150013 using the exponential disk profile from \cite{binneyGalacticDynamicsSecond2008a} and assume a conservative log-uniform prior on $M$ over $[10^{-3}, 10^3] ~M_{\oplus}$. Assuming the lens and source velocities are independently drawn from a 3D Gaussian with a mean of $0$ km/s and standard deviation $\sigma = 55 ~\text{km/s}$, we calculate the $v_T$ by taking the vector difference and projecting onto the transverse plane. This results in a Rayleigh distribution for the transverse velocity $$P(v_T) = \frac{v_T}{\sigma_r^2}\exp\Big(-\frac{v_T^2}{2\sigma_r^2}\Big),$$ where $\sigma_r = \sqrt{\frac{3}{2}} \sigma$ and $v_T \geq 0$.

We use a Gaussian likelihood on $\theta_E$ and $t_E$ with standard deviation given by the mean of their upper and lower uncertainties. Performing the marginalization over this likelihood yields $M = 0.53^{+0.40}_{-0.26} ~M_{\oplus}, d_L = 675^{+359}_{-287} ~\text{pc}, v_T = 63.26^{+31.79}_{-26.46} ~\text{km/s}$ and is not strongly dependent on the assumed priors. For example, adopting a significantly different prior on the mass corresponding to the normalized predicted event rate (cf. Fig. \ref{fig:yields}) originating from the best-fit FFP mass function of \citet{sumi2023freefloating} yields $M = 0.33^{+0.46}_{-0.15} ~M_{\oplus}, d_L = 490^{+425}_{-219} ~\text{pc}, v_T = 43.80^{+40.93}_{-17.91} ~\text{km/s}$.

It is also worth noting that before measuring $\theta_E$ and $t_E$, the predicted event rate alone indicates that TESS's sensitivity to FFPs would peak at masses around $[10^{-2}, 10^{-1}] ~M_{\oplus}$. The recovered mass for the event in question is consistent with this expectation. However, this implicitly assumes that the mass function presented in \citet{sumi2023freefloating} remains a good model of the FFP abundance for $M < M_\oplus$, a mass range that is currently unconstrained by observations. Thus the estimates associated with such a calculation are subject to considerable uncertainty, further complicating the interpretation of this particular event. Taken as a whole, this discussion demonstrates that under the assumption of a range of reasonable priors, the event is consistent with that of a low-mass planet traveling at a reasonable velocity relative to the local stellar distribution \citep{2018A&A...616A..11G}, making it even more challenging to flag as a false positive.

\subsubsection{Consistency with Expected Yield}
\label{subsec:yield}
Due to the expected increased abundance of FFPs with decreasing mass \citep{2023AJ....166..108S,gould_free-floating_2022,mroz2023exoplanet}, any events observed by TESS would be expected to be largely due to FFPs with masses on the order of $10^{-2} - 10^{-1} \, M_\oplus$, below which the typical FFP would not produce a detectable magnification given TESS's sensitivity. This is evident in Figure. \ref{fig:yields}. However, Figure. \ref{fig:yields} also shows that in the \textit{entirety} of existing observations, we expect only on the order of one (fiducial) to twenty ($+1\sigma$) FFPs total.

Given this expected event rate, we can compute the odds of detecting at least one event in a search of five sectors. Summing over all bins in Fig. \ref{fig:yields}, we find that the central power-law predicts $\approx 1/70$ events per sector, averaged over sectors. Assuming detections follow a Poisson process, the probability of detecting at least one event in Sectors 61 -- 65 is given by $1 - P(k = 0, \lambda \approx 1/14) \approx 7\%$, though this ranges from $\approx 0.7\%$ to $\approx 80\%$ for the $\pm 1\sigma$ slope of the mass function. These estimates indicate that the likelihood of detecting an FFP in the TESS data that we have searched is small. This motivates our investigation of alternatives to the lensing interpretation, as discussed in the following section.

\subsection{False Positive Interpretation}
\label{sec:falsepositive}

Although TESS's observing strategy allows for very high-cadence observation of target stars, its per-sector baseline is only 27.4 days. While many targets have been observed in multiple sectors, and hence have significantly longer baselines, these observations are sporadic and not conducive to searches for long-term variability. As such, comparison to external observations made by long-baseline stellar surveys is necessary for disentangling the nature of a signal. OGLE and ASAS-SN data \citep{Shappee2014,Hart2023} described by \citet{Mroz2024} revealed that the star associated with the Sector 61 signal exhibits a low-amplitude variability with a period of $42-44$ days, longer than the TESS baseline. \citet{Mroz2024} suggested that this corresponds to rotational variability due to starspots, which in turn suggests the presence of magnetic activity in the target star. That work concluded that the observed event is likely a stellar flare. 

In light of these observations, we analyze the characteristics of the primary event associated with TIC-107150013 as a stellar flare. As we show below, the apparent morphology of the event, its corresponding flare energy, and its relative isolation would make it a very unusual class of flare, and may point to a pernicious class of false positive that has not been well-characterized in the literature. 

\subsubsection{Consistency with Flare Morphology}\label{sec:morphology}

Flares are expected to be largely asymmetric, with rapid flux increases followed by a slowly-falling tail, a morphology not exhibited in this particular event. We quantified this event's consistency with flare morphology by fitting flare template models described by both \citet{Davenport2014} and \citet{Pitkin2014}. The \citet{Davenport2014} model is commonly used to identify classical flares in dedicated studies of flares, and follows a fourth-order polynomial rise and double exponential decay characterized by the flare peak time ($t_{\mathrm{peak}}$), the peak amplitude ($A$), and the full time width at half the maximum flux ($t_{1/2}$). The \citet{Pitkin2014} model allows for a more flexible flare morphology including more symmetric curves, and involves a rapid rise with a half-Gaussian profile followed by an exponential decay, characterized by the flare peak time, the peak amplitude, the standard deviation of the Gaussian rise ($\tau_{g}$), and the exponential decay constant ($\tau_{e}$). 

\begin{figure*}
\centering
\includegraphics[width=0.49\linewidth]{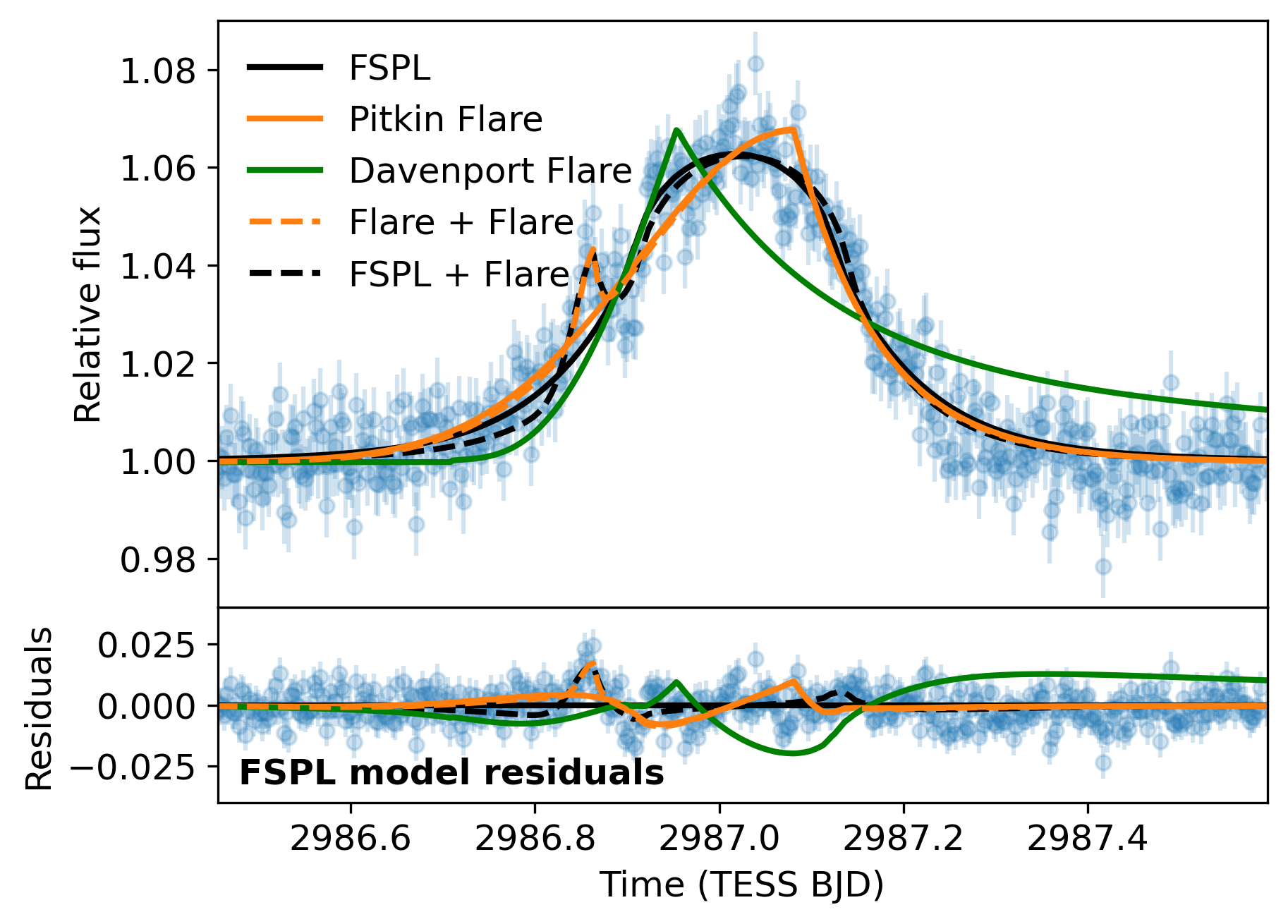}
\includegraphics[width=0.49\linewidth]{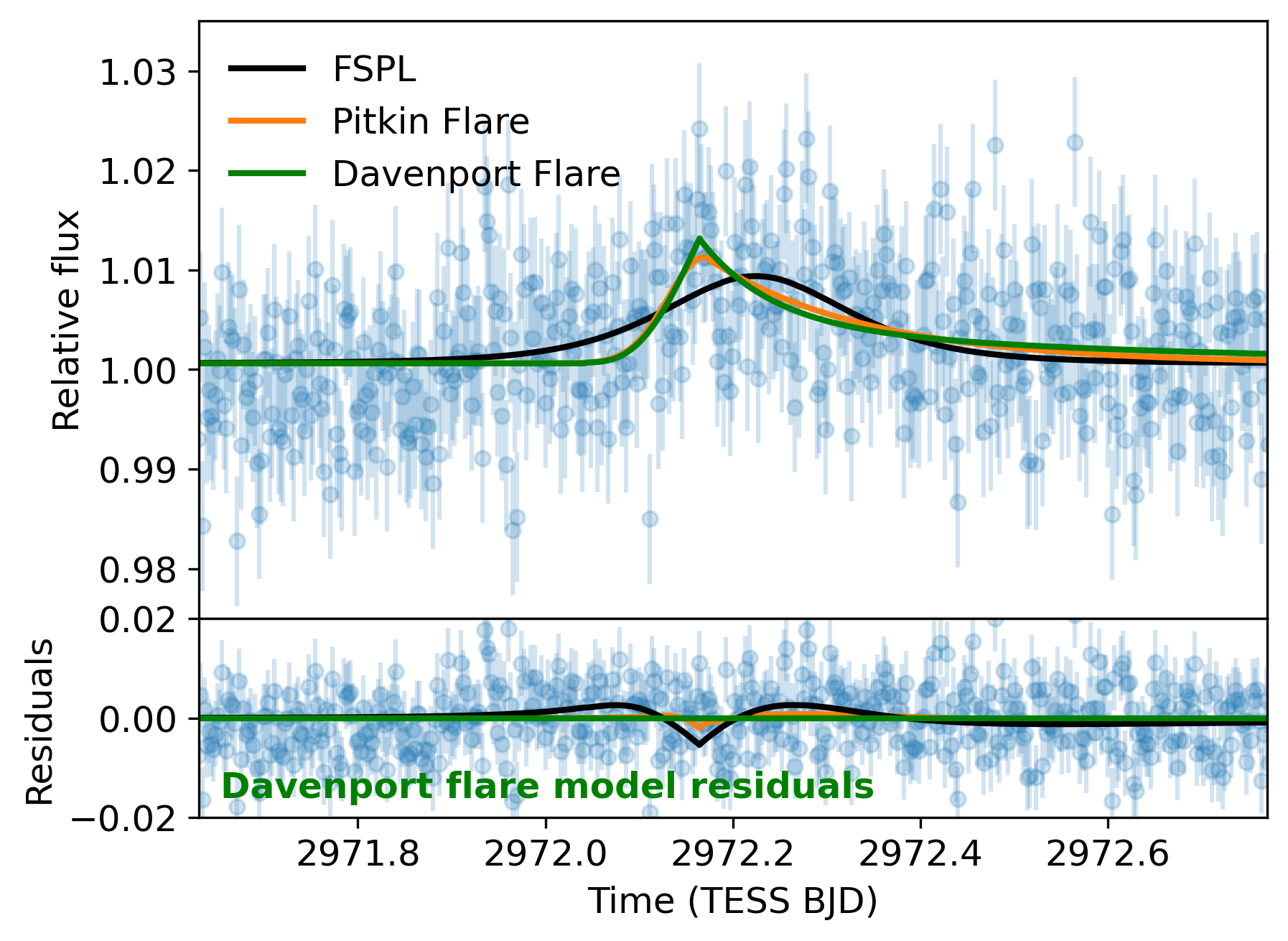}
\caption{The primary brightening event associated with TIC-107150013 (left) and the secondary smaller bump earlier in the light curve (right) overlaid with colored lines showing the best-fit FSPL lensing model (black), \citet{Pitkin2014} flare model (orange), \citet{Davenport2014} flare model (green). We also show a \citet{Pitkin2014} flare + flare model (dotted orange) and FSPL + flare model (dotted black) for the primary event only (see text). The bottom panel shows the residual light curve and each model after removal of the baseline models chosen to interpret the events (FSPL model for the primary event; \citet{Davenport2014} model for the secondary event).}\label{fig:primary}
\end{figure*}

The fit results for all three models are provided in Table \ref{tab:flares} and shown in Figure \ref{fig:primary}. We compare the BIC of these fits to the FSPL model and find $\mathrm{BIC}_{\mathrm{Pitkin}} - \mathrm{BIC}_{\mathrm{FSPL}} = 12.8$ and $\mathrm{BIC}_{\mathrm{Davenport}} - \mathrm{BIC}_{\mathrm{FSPL}} = 1262.9$. In other words, the \citet{Pitkin2014} model is the better flare model, but is still only $\exp{(-\Delta\mathrm{BIC}/2)} = 0.0016$ times as probable as the FSPL model to minimize the information loss. Both models have essentially no support at the level of the data compared to the FSPL model. Furthermore, if we were to impose $\tau_{g} < \tau_{e}$ for the \citet{Pitkin2014} flare model, i.e., we require that the rise is shorter than the subsequent decay (following \citet{Pitkin2014} when using this model for flare detection), then model comparison gives $\mathrm{BIC}_{\mathrm{Pitkin}} - \mathrm{BIC}_{\mathrm{FSPL}} = 98.9$ compared to the FSPL model. Therefore, the best-performing flare model becomes $3\times10^{-22}$ times as probable as the FSPL model to minimize the information loss. These results indicate that existing models of flares struggle to account for events such as the one we have identified, and as such, this event provides a potential example of the new kinds of challenging backgrounds future space-based surveys will face

As mentioned in \S\ref{sec:lensingmodels}, there is a small secondary brightening spike during the rise of the event at $t = 2986.85$ days, particularly visible in the residuals of the TGLC light curve; see left panel of Figure \ref{fig:primary}. We therefore also explore the false positive scenario that the primary event is not one flare, but the summation of two flares. We fit the sum of two \citet{Pitkin2014} flare models, and found $\mathrm{BIC}_{\mathrm{Double}} - \mathrm{BIC}_{\mathrm{FSPL}} = 17.7$ ($64.8$ if we require $\tau_{g} < \tau_{e}$), which still indicates strong preference for the microlensing interpretation at the level of the data. We also fit the sum of an FSPL model and the \citet{Pitkin2014} flare model, and found $\mathrm{BIC}_{\mathrm{FSPL}} - \mathrm{BIC}_{\mathrm{FSPL+Flare}} = 37.5$ ($35.3$ if we require $\tau_{g} < \tau_{e}$), which demonstrates that the combined FSPL and flare model gives the best fit overall, over FSPL alone, a single flare, and a double flare. However, because the FSPL fit parameters are within $1\sigma$ of the results assuming microlensing alone, we do not explore this scenario further in the paper.

While we prefer BIC over AIC because BIC better avoids overfitting for model selection (see \S\ref{sec:lensingmodels}), AIC is a common choice.  When we used AIC for model comparisons we found that the FSPL model alone is still overwhelmingly preferred over both flare models ($\mathrm{AIC}_{\mathrm{Pitkin}} - \mathrm{AIC}_{\mathrm{FSPL}} = 12.8$ and $\mathrm{AIC}_{\mathrm{Davenport}} - \mathrm{AIC}_{\mathrm{FSPL}} = 1270.0$) as well as the double flare model when requiring $\tau_{g} < \tau_{e}$ ($\mathrm{AIC}_{\mathrm{Double}} - \mathrm{AIC}_{\mathrm{FSPL}} = 36.4$). However, the double flare model is preferred over the FSPL model when relaxing this requirement ($\mathrm{AIC}_{\mathrm{FSPL}} - \mathrm{AIC}_{\mathrm{Double}} = 10.6$). The discrepancy between BIC and AIC suggests that the double flare model is overfitting the data. We still find that the combined FSPL and flare model is the overall best performing model according to AIC ($\mathrm{AIC}_{\mathrm{FSPL}} - \mathrm{AIC}_{\mathrm{FSPL+Flare}} = 65.9$), including over the double flare model. Our conclusion that the event is best explained by microlensing at the level of the data remains unchanged.

\begin{figure*}
    \centering
    \includegraphics[width=0.9\linewidth]{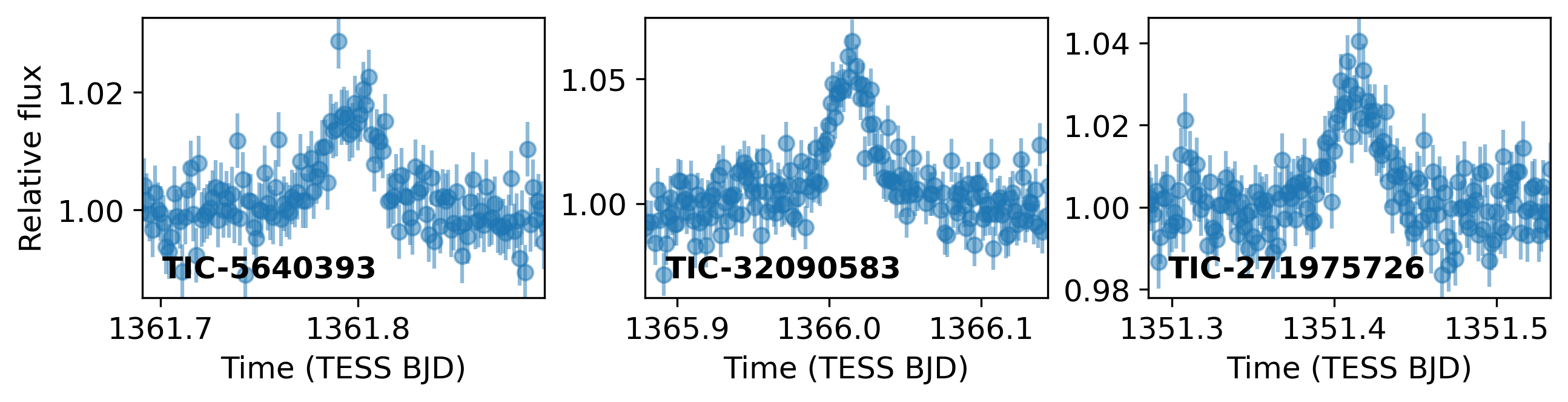}
    \caption{All flares manually flagged as symmetric and microlensing-like out of a subset of 291 flares compiled by \citet{GuntherFlares}. These flares would all have been rejected by our automated vetting tests due to large intrinsic red noise and the presence of other flares in the light curve, but suggest a possible symmetric flare occurrence rate up to $\sim1\%$. See Sec. \ref{sec:falsepositive} for details.}\label{fig:flares}
\end{figure*}

More generally, to our knowledge, no dedicated study of the rate of microlensing-like symmetric flares has been performed. We attempted to estimate the frequency of such events from the catalog of flares compiled by \citet{GuntherFlares} from TESS Sectors 1 and 2 taken at 120-second cadence, as this provides the most direct comparison to our data. We visually inspected 291 flares with properties most similar to microlensing events, based on their timescales (full widths at half-maximum $> 0.03$ days) and peak heights (amplitudes $> 0.5\%$), and flagged  three events as symmetric corresponding to TIC-5640393, 32090583, and 271975726 (Figure \ref{fig:flares}). This could suggest a symmetric flare occurrence rate of $\sim1\%$. However, our automated vetting pipeline failed all three events either due to high levels of red noise resulting in a low event SNR below our detection threshold, or other strong flares in the light curve resulting in the event being considered non-unique, and FSPL fits to these events failed to converge. Nevertheless, using the maximum likelihood results after 50,000 steps, we found that model comparison was unable to distinguish between interpretations to the same significance as TIC-107150013. TIC-5640393 corresponds to $\mathrm{BIC}_{\mathrm{Pitkin}} - \mathrm{BIC}_{\mathrm{FSPL}} = 0.1$ while TIC-32090583 corresponds to $\mathrm{BIC}_{\mathrm{FSPL}} - \mathrm{BIC}_{\mathrm{Pitkin}} = 0.1$, indicating no evidence for preferring either the flare or FSPL model \cite[$\Delta\mathrm{BIC} < 2$;][]{Kass01061995}. There is evidence for favoring the FSPL model over the best flare model for TIC-271975726 ($\mathrm{BIC}_{\mathrm{Pitkin}} - \mathrm{BIC}_{\mathrm{FSPL}} = 2.6$), still significantly weaker evidence than what we found for TIC-107150013 ($\mathrm{BIC}_{\mathrm{Pitkin}} - \mathrm{BIC}_{\mathrm{FSPL}} = 12.8$). Combined with the fact that these flares failed automated vetting and that their FSPL fits failed to converge, we conclude that the primary event associated with TIC-107150013 is unique compared to these other symmetric events based on morphology.

\begin{table*}
    \centering
    \caption{Best-fit model results for the FSPL lensing model and \citet{Pitkin2014} and \citet{Davenport2014} flare models, for the primary event associated with TIC-107150013, a smaller bump earlier in the same light curve which could be a flare, and the three symmetric flares identified from the \citet{GuntherFlares} flare catalog (\S\ref{sec:morphology}). BIC values are included for model comparison. Because the FSPL fits only converged for the TIC-107150013 primary event and secondary bump. The FSPL properties for the symmetric flares were obtained after running \texttt{emcee} for 50,000 steps (2.5 million samples) which was not sufficient for convergence, and should therefore not be considered robust FSPL parameters.}\label{tab:flares}
    \begin{tabular}{lccccc}
        \hline\hline
         Parameter & TIC-107150013 (primary) & TIC-107150013 (bump) & TIC-5640393 & TIC-32090583 & TIC-271975726 \\ 
        \hline\hline
        \multicolumn{6}{c}{\textit{FSPL lensing model}}\\
        \hline
        $t_{0}$ (BJD - 2457000) & $2987.017_{-0.002}^{+0.002}$ & $2972.224_{-0.013}^{+0.015}$ & $1361.797_{-0.001}^{+0.001}$ & $1366.014_{-0.001}^{+0.001}$ & $1351.414_{-0.001}^{+0.001}$ \\
        $u_{0}$ & $4.33_{-0.15}^{+0.15}$ & $5.32_{-1.53}^{+2.78}$ & $8.41_{-4.61}^{+0.70}$ & $2.62_{-0.48}^{+1.23}$ & $4.94_{-2.22}^{+2.01}$  \\
        $t_{E}$ (days) & $0.074_{-0.002}^{+0.002}$ & $0.043_{-0.006}^{+0.006}$ & $0.005_{-0.001}^{+0.001}$ & $0.009_{-0.001}^{+0.001}$ & $0.007_{-0.002}^{+0.001}$ \\
        $\rho$ & $4.57_{-0.15}^{+0.15}$ & $4.78_{-3.25}^{+3.75}$ & $8.73_{-6.02}^{+0.92}$ & $1.9_{-1.3}^{+2.1}$ & $5.14_{-3.54}^{+2.20}$ \\
        $f_\mathrm{source} ~(e^{-}s^{-1})$ & $875_{-52}^{+52}$ & $865_{-53}^{+53}$ & $150_{-7}^{+7}\times10^{6}$ & $701_{-59}^{+59}$ & $942_{-56}^{+60}$\\
        $f_\mathrm{blend} ~(e^{-}s^{-1})$ & $-11_{-52}^{+52}$ & $0_{-53}^{+53}$ & $1_{-7}^{+7}\times10^{6}$ & $0_{-61}^{+57}$ & $-3_{-58}^{+58}$ \\
        BIC & $56232.9$ & $65048.4$ & $758612.0$ & $120883.6$ & $166155.7$ \\
        \hline
        \multicolumn{6}{c}{\textit{\citet{Davenport2014} flare model}}\\
        \hline
        $t_{\mathrm{peak}}$ (BJD - 2457000) & $2986.954_{-0.002}^{+0.002}$ & $2972.164_{-0.006}^{0.005}$ & $1361.790_{-0.001}^{+0.001}$ & $1366.007_{-0.001}^{+0.001}$ & $1351.407_{-0.001}^{+0.001}$ \\
        $t_{1/2}$ (days) & $0.243_{-0.006}^{+0.006}$ & $0.116_{-0.023}^{+0.017}$& $0.013_{-0.002}^{+0.002}$ & $0.022_{-0.002}^{+0.001}$ & $0.019_{-0.002}^{+0.002}$ \\ 
        $A$ & $0.069_{-0.004}^{+0.005}$ & $0.013_{-0.002}^{+0.002}$ & $0.024_{-0.003}^{+0.003}$ & $0.063_{-0.006}^{+0.007}$ &  $0.035_{-0.003}^{+0.004}$ \\
        $f_\mathrm{source} ~(e^{-}s^{-1})$ &  $860_{-55}^{+56}$ & $860_{-54}^{+54}$ & $150_{-7}^{+7}\times10^{6}$ & $694_{-61}^{+61}$ & $937_{-61}^{+61}$\\
        $f_\mathrm{blend} ~(e^{-}s^{-1})$ & $4_{-56}^{+55}$ & $4_{-54}^{+54}$ & $0_{-7}^{+7}\times10^{6}$ & $4_{-61}^{+61}$ & $4_{-61}^{+61}$ \\
        BIC & $57495.8$ & $65035.5$ & $758641.0$ & $120943.4$ & $166194.2$ \\
        \hline
        \multicolumn{6}{c}{\textit{\citet{Pitkin2014} flare model}}\\
        \hline
        $t_{\mathrm{peak}}$ (BJD - 2457000) & $2987.081_{-0.005}^{+0.004}$ & $2972.194_{-0.027}^{+0.042}$ & $1361.805_{-0.002}^{+0.002}$ & $1366.017_{-0.001}^{+0.001}$ &  $1351.418_{-0.002}^{+0.002}$\\
        $\tau_{g}$ (days) &  $0.170_{-0.005}^{+0.005}$ & $0.067_{-0.03}^{+0.04}$ & $0.019_{-0.002}^{+0.002}$ & $0.015_{-0.002}^{+0.002}$ & $0.016_{-0.002}^{+0.002}$ \\ 
        $\tau_{e}$ (days) &  $0.091_{-0.004}^{+0.005}$ & $0.165_{-0.031}^{+0.032}$ &$0.018_{-0.002}^{+0.003}$ &$0.016_{-0.002}^{+0.002}$ & $0.013_{-0.002}^{+0.003}$\\
        $A$ & $0.068_{-0.004}^{+0.005}$ & $0.07_{-0.05}^{+0.22}$ & $0.018_{-0.002}^{+0.002}$ & $0.058_{-0.006}^{+0.006}$ & $0.031_{-0.003}^{+0.003}$ \\
        $f_\mathrm{source} ~(e^{-}s^{-1})$ & $860_{-56}^{+55}$ & $128_{-98}^{+413}$& $150_{-7}^{+7}\times10^{6}$ & $694_{-61}^{+62}$ & $935_{-61}^{+60}$\\
        $f_\mathrm{blend} ~(e^{-}s^{-1})$ & $4_{-55}^{+56}$ & $736_{-413}^{+98}$ & $0_{-7}^{+7}\times10^{6}$ & $4_{-61}^{+61}$ & $6_{-60}^{+61}$ \\
        BIC & $56245.8$ & $65042.0$ & $758612.1$ & $120883.6$ & $166158.4$ \\
\end{tabular}
\end{table*}

Finally, there is a smaller, flare-like brightening feature in the light curve for TIC-107150013 near 2972 days (Figure \ref{fig:primary}), which could indicate potential activity on the star. Both flare models of this event are preferred over the FSPL model  ($\mathrm{BIC}_{\mathrm{FSPL}} - \mathrm{BIC}_{\mathrm{Davenport}} = 12.9$, $\mathrm{BIC}_{\mathrm{FSPL}} - \mathrm{BIC}_{\mathrm{Pitkin}} = 6.4$; Table \ref{tab:results}). However, this event has a low significance ($\mathrm{SNR} = 6.5$), below our detection threshold, and fails our Uniqueness Test (\S\ref{sec:unique}) even after masking out the primary event at 2987 days. We also found that the light curve for TIC-1017150013 features significant amounts of red noise at the timescale of this event (0.32 days), where the amount of red noise ($\sigma_{r} = 0.00096$) exceeds the amount of white noise ($\sigma_{w}/\sqrt{n} = 0.00054$). A high amount of red noise is expected for high-precision light curves meant for transit searches \citep{Pont2006}. Combined with the low significance and lack of uniqueness, we cannot rule out this event as correlated noise.

\subsubsection{Consistency with Flare Energies}

\begin{figure}
\centering
\includegraphics[width=\linewidth]{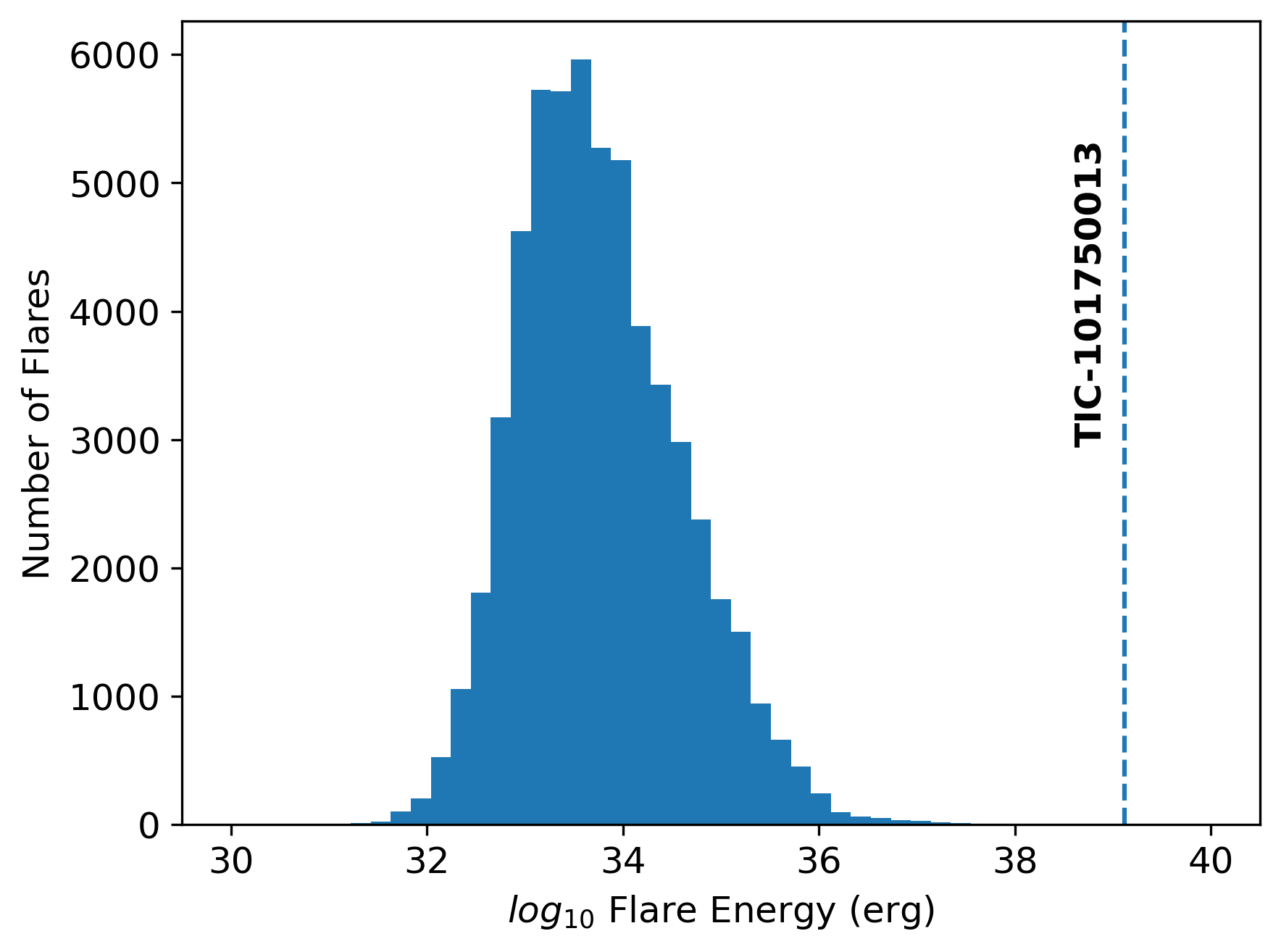}
\caption{Histogram of flare energies for $57,961$ flares measured by \citet{Yang2023} from TESS sectors 1 -- 30, compared to the measured flare energy of the primary brightening event associated with TIC-107150013.}\label{fig:flare_energies}
\end{figure}

We also computed the expected bolometric flare energy of the event, $E_{\mathrm{flare}}$, to assess its consistency with typical flares. Following the procedure described by \citet{Gunther2020}, the luminosities of the star and flare in the TESS bandpass assuming black-body radiation are given by
\begin{equation}
L_{\star}^{\prime} = \pi R_{\star}^{2}\int R_{\lambda}B_{\lambda}(T_{\mathrm{eff}}) d\lambda,
\end{equation}
\begin{equation}
L_{\mathrm{flare}}^{\prime}(t) = A_{\mathrm{flare}}(t)\int R_{\lambda}B_{\lambda}(T_{\mathrm{flare}}) d\lambda,
\end{equation}
where $R_{\lambda}$ is the TESS response function,\footnote{\url{https://heasarc.gsfc.nasa.gov/docs/tess/data/tess-response-function-v2.0.csv}} $B_{\lambda}(T)$ is the Planck function evaluated at a given temperature $T$, and $A_{\mathrm{flare}}(t)$ is the time-dependent area of the flare. We solve for $A_{\mathrm{flare}}$ by recognizing that the normalized light curve gives the relative flare amplitude $(\Delta F/F)(t) = L_{\mathrm{flare}}^{\prime}(t)/L_{\star}$, giving
\begin{equation}
A_{\mathrm{flare}}(t) = (\Delta F/F)(t) \pi R_{\star}^{2} \frac{\int R_{\lambda}B_{\lambda}(T_{\mathrm{eff}}) d\lambda}{\int R_{\lambda}B_{\lambda}(T_{\mathrm{flare}}) d\lambda}.
\end{equation}
We then find the bolometric flare luminosity $L_{\mathrm{flare}}(t)$ as
\begin{equation}
L_{\mathrm{flare}}(t) = \sigma_{\mathrm{SB}}T_{\mathrm{flare}}^{4}A_{\mathrm{flare}}(t),
\end{equation}
and integrate over time to get the bolometric energy of the flare,
\begin{equation}\label{eqn:Eflare}
E_{\mathrm{flare}} = \int L_{\mathrm{flare}}(t) dt.
\end{equation}
We take $T_{\mathrm{eff}} = 4115$ K and $T_{\mathrm{flare}} = 9000$ K for the star and flare, respectively, where 9000 K is a conservative lower limit consistent with previous studies \cite[e.g.][]{Shibayama2013,GuntherFlares}. We model $(\Delta F/F)(t)$ using the best-fit \citet{Pitkin2014} flare model. Integrating Eqn. \ref{eqn:Eflare} over the event, we find $E_{\mathrm{flare}} = 1.3\times10^{39}~\mathrm{erg}$, indicating that the event could constitute some form of  unusually energetic superflare, which have typical energies of $10^{33} - 10^{38}~\mathrm{erg}$ \cite[e.g.][]{Shibayama2013}. The energy release returned by the above estimate is at least two orders of magnitude more energetic than all 8695 flares observed at 2 minute cadence in the first two sectors of TESS data \citep{GuntherFlares}, the most energetic of which had $E_{\mathrm{flare}} = 10^{36.9}~\mathrm{erg}$, which corresponded to the G-type giant star TIC-332487879. It would also be more energetic than all 57,961 flares with energies measured from TESS sectors 1 -- 30 by \citet{Yang2023}, of which only 10 (0.017\%) had $E_{\mathrm{flare}} > 10^{38}~\mathrm{erg}$ (Figure \ref{fig:flare_energies}).

The flare energies associated with the symmetric events around TIC-32090583 and 271975726 are also consistent with expectations for flares \citep{Shibayama2013, GuntherFlares, Yang2023}, with $E_{\mathrm{flare}} = 6.8\times10^{34}~\mathrm{erg}$ and $4.7\times10^{34}~\mathrm{erg}$, respectively. We cannot robustly estimate $E_{\mathrm{flare}}$ for TIC-5640393 because it lacks known stellar properties; however, assuming a Sun-like star we find $E_{\mathrm{flare}} = 2.1\times10^{35}~\mathrm{erg}$, and assuming a TIC-107150013-like star we find $E_{\mathrm{flare}} = 3.6\times10^{37}~\mathrm{erg}$, both of which are still in line with expectations for flare energies.

If the event identified on TIC-107150013 is interpreted as a flare, it is significantly more energetic than other observed flares and is poorly described by existing flare models. The event is unique compared to other symmetric events from \citet{GuntherFlares} not only in terms of morphology, but also in terms of flare energy. While such a flare is possible, its lack of precedent in existing observations imply that it is very rare. We conclude that the flare false positive interpretation is unlikely, and, given the lack of an observational dataset of similar flares, we cannot quantitatively estimate how unlikely it is relative to the FFP interpretation.

\subsubsection{Non-Flare False Positive Explanations}

Other false positive scenarios are possible but are similarly challenged. For example, an event with a morphology similar to our signal has been observed on a young M-dwarf with an extremely strong ($>1$ KG) magnetosphere by \citep{Palumbo2022}.  They interpret this event as the centrifugal breakout of a clump of stellar material that has been entrapped in a very strong rigidly rotating magnetosphere (RRM).  Evidence for RRMs have been observed on hot A and B stars \citep{Townsend2005,Oksala2012}, and on very young M-dwarfs \citep{Palumbo2022}. However, RRMs require a strong ($>1$ KG) globally organized and stable magnetic field \citep{Palumbo2022}. The stellar properties of TIC-107150013 in Table~\ref{tab:star} indicate that it is a cool, large, and therefore evolved, star which is not expected to have such a strong magnetosphere, hence this too remains an unlikely scenario.

TIC-107150013 appears to be a variable star based on OGLE data so, following common practice, we would not recommend TIC-107150013 to be considered in a general search for FFPs. Despite this, the signal we observe on TIC-107150013 provides an example of an FFP-like event arising from some rare astrophysical phenomenon that has not been thoroughly explored in the literature. Since TESS's targets are nearby and their properties are well-characterized, our ongoing search through other TESS sectors will provide an opportunity to better understand such events and correlate them with the properties of the stars on which they appear, an important resource for future microlensing efforts.

\section{Summary and Future Directions}

The search presented here is preliminary, having been based on a small and relatively bright subset of all TESS target stars. We are currently repeating our search across all TESS sectors, and plan to use a fainter dataset of TGLC light curves down to $T = 16$ mag. This combination will increase the number of possible source stars by over two orders of magnitude. These data will provide the opportunity to continue searching for FFP signals in the sub-terrestrial mass range. Furthermore, our ongoing search for such signals will provide key insight on possible sources of false positive contamination for future space-based missions such as the Roman Space Telescope's Galactic Bulge Time Domain Survey (GBTDS) \citep{Akeson2019, Penny2019} and Earth 2.0 \citep{Ge2022}.

Our initial results presented here demonstrate that TESS can be a powerful tool to search for signals like that of free-floating planets at masses below that of Earth. The results of our ongoing search will play a key role in calibrating the expected yield and search strategy of the GBTDS. Though never intended as a microlensing satellite, TESS is helping to open a window to the dark, enigmatic population of worlds that lie drifting between the stars.

\begin{acknowledgments}
We thank the referee for their useful feedback which significantly improved the manuscript. We thank  J.P. Beaulieu for his useful input on revising this manuscript, Marc Hon for discussions about giant stars and activity, Te Han for discussions about TGLC light curves, Avi Shporer for discussions about follow-up prospects for our events, and Jon Jenkins for discussions about SPOC 2-min and 20-sec cadence light curves. MK acknowledges the support of the Natural Sciences and Engineering Research Council of Canada (NSERC), RGPIN-2024-06452. Cette recherche a été financée par le Conseil de recherches en sciences naturelles et en génie du Canada (CRSNG), RGPIN-2024-06452. MK also acknowledges support by the Juan Carlos Torres postdoctoral fellowship from the MIT Kavli Institute for Astrophysics and Space Research. WD was supported by NSF grant PHY-2210361 and the Maryland Center for Fundamental Physics. WD and NS also acknowledge the support of DOE grant No. DE-SC0010107. BSG was supported by NASA Grant 80NSSC24M0022 and by the Thomas Jefferson Chair Endowment for Discovery and Space Exploration.

This paper includes data collected by the TESS mission, which are publicly available from the Mikulski Archive for Space Telescopes (MAST). Funding for the TESS mission is provided by NASA’s Science Mission directorate. We acknowledge the use of public TESS data from pipelines at the TESS Science Office and at the TESS Science Processing Operations Center. This work has made use of data from the European Space Agency (ESA) mission {\it Gaia} (\url{https://www.cosmos.esa.int/gaia}), processed by the {\it Gaia} Data Processing and Analysis Consortium (DPAC, \url{https://www.cosmos.esa.int/web/gaia/dpac/consortium}). Funding for the DPAC has been provided by national institutions, in particular the institutions participating in the {\it Gaia} Multilateral Agreement.
\end{acknowledgments}

%

\vspace{5mm}
\facilities{Gaia, TESS}


\software{\texttt{lmfit} \citep{LMFIT}, \texttt{matplotlib} \citep{Hunter2007}, \texttt{numpy} \citep{Harris2020}, \texttt{pandas} \citep{reback2020pandas, mckinney-proc-scipy-2010}, \texttt{pyLIMA} \citep{pylima}, \texttt{scipy} \citep{Virtanen2020}, \texttt{tess-point} \citep{tesspoint}}





\bibliography{sample701}{}
\bibliographystyle{aasjournal}



\end{document}